\documentclass[iop]{emulateapj}
\usepackage{natbib}
\shorttitle{Escape zones of charged particles around magnetized black hole}
\shortauthors{Kop\'{a}\v{c}ek \& Karas}

\begin{document}
\title{Near-horizon structure of escape zones of electrically charged particles around weakly magnetized rotating black hole}
\email{kopacek@ig.cas.cz}
\author{Ond\v{r}ej Kop\'{a}\v{c}ek}
\author{Vladim\'{i}r Karas}
\affiliation{Astronomical Institute, Czech Academy of Sciences, Bo\v{c}n\'{i} II, CZ-141\,31~Prague, Czech~Republic.}

\begin{abstract}
An interplay of magnetic fields and gravitation drives accretion and outflows near black holes. However, a specific mechanism is still a matter of debate; it is very likely that different processes dominate under various conditions. In particular, for the acceleration of particles and their collimation in jets, an ordered component of the magnetic field seems to be essential. Here we discuss the role of large-scale magnetic fields in transporting the charged particles and dust grains from the bound orbits in the equatorial plane of a rotating (Kerr) black hole and the resulting acceleration along trajectories escaping the system in a direction parallel to the symmetry axis (perpendicular to the accretion disk). We consider a specific scenario of destabilization of circular geodesics of initially neutral matter by charging (e.g., due to photoionization). Some particles may be set on escaping trajectories and attain relativistic velocity. The case of charged particles differs from charged dust grains by their charge-to-mass ratio, but the acceleration mechanism operates in a similar manner. It appears that the chaotic dynamics controls the outflow and supports the formation of near-horizon escape zones.  We employ the technique of recurrence plots to characterize the onset of chaos in the outflowing medium. We investigate the system numerically and construct the basin-boundary plots, which show the location and the extent of the escape zones. The effects of black hole spin and magnetic field strength on the formation and location of escape zones are discussed, and the maximal escape velocity  is computed.
\end{abstract}


\section{Introduction}
\label{intro}
Since the seminal works of \citet{blandford77} and \citet{blandford82}, rotation of the black hole with the surrounding accretion disk and a large-scale magnetic field are considered as key ingredients for the effective acceleration and collimation of outflows of matter from the vicinity of accreting black holes. The Blandford--Znajek process operates near the event horizon owing to the frame-dragging effect, which amplifies and twists the magnetic field anchored in the accretion disk. Similarly, the Blandford--Payne process requires a magnetized accretion disk. As a result, the charged matter of the disk may be accelerated outward in jets aligned with the spin axis, transporting away the rotational energy of the black hole.  On the other hand, small-scale magnetic fields may be induced in the accretion disk owing to turbulence driven by the magnetorotational instability \citep{balbus98}. While turbulence is effective in the transport of the angular momentum (i.e., it contributes to the viscosity of the disk), the resulting small-scale field structure seems insufficient for launching the jet \citep{beckwith08,gold17}. 

The crucial role of the ordered large-scale magnetic fields in powering the outflows (and astrophysical jets, in particular) is being confirmed by numerical experiments. Indeed, in 3D general relativistic megnetohydrodynamic (GRMHD) simulations of accretion processes, ordered bundles of magnetic field lines seem to emerge in a self-consistent manner under rather general conditions \citep[e.g.,][]{penna10,tchekhovskoy15,sadowski16}. The concept of the Blandford--Znajek or Blandford--Payne process is thus further supported as a plausible mechanism of launching the outflow. 

In the given context we investigate the motion of charged particles (e.g., heavy ions, dust grains) near the magnetized rotating black hole. In particular, we discuss under which circumstances the material bound to stable orbits embedded in the equatorial plane may be set on the escaping trajectories, liberated from the attraction of the center and even accelerated to relativistic velocities. We consider a single particle limit, i.e. the particles are considered as noninteracting and hydrodynamical terms are neglected in governing equations. Direct applicability of the model to astrophysical systems is thus limited to the regions where the gas is sufficiently diluted (the mean free path of particles is larger than the characteristic length scale of the system given by the radius of event horizon). Nevertheless, the main motivation of this study lies in conceptual questions concerning the effect of the magnetic field on the stability of orbits near the strongly gravitating center. 

Previously, we have demonstrated that the structure of the electromagnetic field is strongly influenced by the relativistic effects of a spinning compact object \citep{karas09, karas14, karas12}. In particular, we have shown that close to the event horizon, where the frame dragging dominates, these effects shape the field lines (described by the vacuum solution of Maxwell's equations) in a way that may mimic the behavior typically associated with magnetohydrodynamic plasma. Namely, we observed that the structure of the field in some regions of spacetime may correspond to current sheets or X-type null points of magnetic reconnection. As the field's topology changes profoundly owing to gravity, also the motion of charged matter is strongly affected. As a  most striking consequence, the particle dynamics undergoes a transition from regularity to deterministic chaos even in the axisymmetric case of uniform magnetic field aligned along the rotation axis \citep{kopacek10,kopacek10b}. Moreover, we have seen \citep{kopacek14} that any deviation from the perfect axisymmetry leads to overall dominance of chaotic dynamics in the given setup. 

According to general relativity, the energy density of 
the electromagnetic field (as well as any other field) stands on the 
right-hand side in Einstein's field equations for the spacetime structure. 
Hence, the electromagnetic field contributes to the metric; to derive a 
self-consistent solution, the coupled Einstein and Maxwell field 
equations must be solved simultaneously. However, the energy density in astrophysically realistic electromagnetic 
fields (proportional to the square of intensities, $\propto E^2+B^2$) is 
far too low to influence the spacetime noticeably. In these 
circumstances, test-field (linearized in $E$ and $B$ terms) solutions 
are adequate and accurate enough for describing the motion of plasma and 
particles, while the corresponding exact solutions of the Einstein--Maxwell set of equations are mainly of theoretical rather than astrophysical interest.

In particular, in the present discussion we adopt the axisymmetric test-field solution corresponding to an asymptotically uniform magnetic field aligned with the rotation axis \citep{wald74}. This assumption seems astrophysically relevant, and also GRMHD simulations reveal a ``magneto-spin alignment" mechanism causing magnetized disks and jets to align with the spin near black holes \citep{mckinney13}. We have seen \citep{kopacek10} that in this setup regular and chaotic trajectories of charged particles generally coexist since the magnetic field acts as a nonintegrable perturbation. Besides triggering chaos, the field also changes the topology of the effective potential in such a way that it in principle allows the particle to leave the equatorial plane and escape to the asymptotic region in the narrow corridor aligned with the symmetry axis \citep[see Figs. 12-14 in][]{kopacek10}. In this paper we shall investigate this phenomenon systematically and also make some astrophysically motivated assumptions considering the initial conditions of the trajectories. Namely, we focus on initially neutral particles on equatorial co-rotating circular (Keplerian) orbits. Below the innermost stable circular orbit (ISCO), the particle is supposed to freely fall onto the horizon. At some radius $r_0$ (below or above the ISCO), the particle obtains electric charge (by photoionization or other process), and its dynamics changes owing to magnetic field. Within this scenario we study under which circumstances the escaping trajectories may be realized and discuss the maximal acceleration that can be achieved. 
 
Charged particles escaping the attraction of black holes immersed in magnetic field have already been the subject of several studies. Weakly magnetized Schwarzschild black holes have been considered by \citet{frolov10} and \citet{alzahrani13}. An exact solution describing the static black hole immersed in asymptotically uniform magnetic field (Ernst's spacetime) was employed by \citet{huang15}. The case of a spinning black hole with a magnetic test field has also been analyzed \citep{alzahrani14,shiose14}. The special case of a slowly rotating black hole was considered by \citet{hussain14}. Moreover, trajectories of particles escaping from the vicinity of magnetized naked singularities were also examined \citep{babar16}. In these studies, the initial setup consists of circular equatorial orbits of charged particles that are kicked from the equatorial plane mechanically by the collision with another particle or photon. The immediate cause of destabilization of a circular orbit that may lead to escaping trajectory is thus an ad hoc introduction of velocity component perpendicular to the equatorial plane. In our scenario, however, no such kick is needed, as we destabilize the equatorial geodesic orbit of initially neutral particles by the charging process itself. Analogical setup was recently considered by \citet{stuchlik16} with the class of spherical orbits \citep{wilkins72}, and the authors demonstrated that charging the neutral particle on such an orbit (with a nonzero velocity component in the direction normal to the equatorial plane) may initiate its escape. Here we discuss in detail the effect of ionization of particles on a circular (or inspiraling below ISCO) equatorial orbits, which have no obvious tendency to leave the equatorial plane, and thus their escape seems less probable compared to the above-referenced models.
  
The paper is organized as follows. In Section~\ref{spec} we describe the employed model of the weakly magnetized rotating black hole and review the equations of motion for charged particles. Trajectories of particles escaping from the circular orbits in the equatorial plane are discussed in Section~\ref{outflow}. Using the effective potential method, we first identify the necessary conditions for the escape analytically in Section~\ref{ionization}. In the following we numerically integrate the trajectories and discuss the role of the black hole's spin and magnetic field in the emergence of escaping orbits (Section~\ref{escape_zone}). Acceleration of escaping particles is investigated quantitatively in Section~\ref{acceleration}. Results are briefly discussed in Section~\ref{discussion} and summarized in Section~\ref{conclusions}.

\section{Specification of the model, equations of motion}
\label{spec}
Kerr metric  describing the geometry of the spacetime around the rotating black hole may be expressed in Boyer--Lindquist coordinates $x^{\mu}= (t,\:r, \:\theta,\:\varphi)$ as follows \citep{mtw}:
\begin{eqnarray}
\label{metric}
{\rm d}s^2&=&-\frac{\Delta}{\Sigma}\Big({\rm d}t-a\sin{\theta}\,{\rm d}\varphi\Big)^2\\& &\nonumber+\frac{\sin^2{\theta}}{\Sigma}\Big[\big(r^2+a^2)\,{\rm d}\varphi-a\,{\rm d}t\Big]^2+\frac{\Sigma}{\Delta}\,{\rm d}r^2+\Sigma\, {\rm d}\theta^2,
\end{eqnarray}
where
\begin{equation}
{\Delta}\equiv{}r^2-2Mr+a^2,\;\;\;
\Sigma\equiv{}r^2+a^2\cos^2\theta.
\end{equation}
The coordinate singularity at $\Delta=0$ corresponds to the outer/inner horizon of the black hole, $r_\pm=M\pm\sqrt{M^2-a^2}$. Rotation of the black hole is measured by the spin parameter $a\in\left<-M,M\right>$. Here we only consider $a\geq0$ without the loss of generality.

We stress that geometrized units are used throughout the paper. Values of basic constants (gravitational constant $G$, speed of light $c$, Boltzmann constant $k$, and Coulomb constant $k_C$) therefore equal unity, $G=c=k=k_C=1$.

We consider the presence of an organized magnetic field threading the black hole horizon. Black holes do not support their own magnetic field of intrinsic origin, like, e.g., stars can. In order to endow a rotating black hole with a weak, large-scale magnetic component, we employ the test-field solution \citep{wald74} for an asymptotically uniform magnetic field aligned with the rotation axis (i.e., the system remains axially symmetric and stationary). Near the black hole, the field becomes twisted owing to the frame-dragging effect. Moreover, an electric field component is induced here. The vector potential can be expressed in terms of Kerr metric coefficients
(\ref{metric}) as follows:
\begin{eqnarray}
\label{waldpot1}
A_t=\textstyle{\frac{1}{2}}B\left(g_{t\phi}+2a\,g_{tt}\right),\;A_{\phi}=\textstyle{\frac{1}{2}}B\left(g_{\phi\phi}+{2a}g_{t\phi}\right),
\end{eqnarray}
where $B$ is magnetic intensity in the asymptotic region.

Equations of motion for charged particles of charge $q$ and rest mass $m$ may be derived using Hamiltonian $\mathcal{H}$, defined as
\begin{equation}
\label{hamiltonian}
\mathcal{H}=\textstyle{\frac{1}{2}}g^{\mu\nu}(\pi_{\mu}-qA_{\mu})(\pi_{\nu}-qA_{\nu}),
\end{equation}
where $\pi_{\mu}$ is the generalized (canonical) momentum and $g^{\mu\nu}$ denotes contravariant components of the metric tensor. The equations of motion are given as
\begin{equation}
\label{hameq}
\frac{{\rm d}x^{\mu}}{{\rm d}\lambda}\equiv p^{\mu}=
\frac{\partial \mathcal{H}}{\partial \pi_{\mu}},
\quad 
\frac{d\pi_{\mu}}{d\lambda}=-\frac{\partial\mathcal{H}}{\partial x^{\mu}},
\end{equation}
where $\lambda\equiv\tau/m$ is the affine parameter, which becomes dimensionless in geometrized units ($\tau$ denotes the
proper time). Kinematical four-momentum $p^{\mu}$ is given as $p^{\mu}=\pi^{\mu}-qA^{\mu}$, which follows directly from the first equation (\ref{hameq}). The conserved value of the Hamiltonian is therefore given as $\mathcal{H}=-m^2/2$. Moreover, the system is stationary,  and momentum $\pi_t$ is thus an integral of motion expressing (negatively taken) energy of the test particle $\pi_t\equiv-E$. Axial symmetry of the system ensures that the corresponding component of the angular momentum also remains conserved as another integral of motion, $\pi_\varphi\equiv L$. Nevertheless, in the following we switch to specific quantities $E/m\rightarrow E$, $L/m\rightarrow L$, and $q/m\rightarrow q$ when describing the test particle.

The effective potential expressing minimal allowed energy of charged test particles in axisymmetric stationary spacetime  may be derived as follows \citep{kopacek10}:

\begin{equation}
\label{eff}
V_{\rm eff}(r,\theta)=\left(-\beta+\sqrt{\beta^2-4\alpha\gamma}\right)/2\alpha,
\end{equation}
where
\begin{eqnarray}
\label{coeff}
\alpha=&-&g^{tt},\;\;\;\;\;\beta=2\left[g^{t\varphi{}}(L-qA_{\varphi})-g^{tt}qA_{t}\right],\\
\gamma=&-&g^{\varphi\varphi}(L-qA_{\varphi})^2+2g^{t\varphi{}}qA_{t}(L-qA_{\varphi})\\
\nonumber &-&g^{tt}q^2A_{t}^2-m^2.
\end{eqnarray}
Analysis of the two-dimensional effective potential (\ref{eff}) may be used to localize circular orbits not only in the equatorial plane \citep{kolos15} but also off this plane \citep{kovar08,kovar10, kovar13}, and it allows us to discuss the stability of resulting orbits \citep{tursunov16}.
\newpage
\section{Escape of electrically charged particles from the magnetized accretion disk}
\label{outflow}
\subsection{Charging the Matter and Necessary Conditions for the Escape}
\label{ionization}
We consider a standard (thin-disk) accretion geometry \citep{novikov73} in which the electrically neutral matter gradually sinks between co-rotating (prograde, equatorial) Keplerian orbits whose energy $E_{\rm Kep}(r)$ and angular momentum $L_{\rm Kep}(r)$ are given as follows \citep{bardeen72}:
\begin{eqnarray}
 \label{kepconst}
E_{\rm Kep}&=&\frac{r^2-2Mr+ a \sqrt{Mr}}{r\sqrt{r^2-3Mr+ 2a\sqrt{Mr}}},\\
L_{\rm Kep}&=&\frac{\sqrt{M} (r^2+a^2- 2a\sqrt{Mr})}{\sqrt{r(r^2-3Mr+ 2a\sqrt{Mr})}}.
\end{eqnarray}
Co-rotating circular geodesics are not allowed below the radius of marginally stable orbit $r_{\rm ms}$ whose position is given as
\begin{equation}
\label{rms}
r_{\rm{ms}}=M\left(3+Z_2-\sqrt{(3-Z_1)(3+Z_1+2Z_2)}\right),
\end{equation} 
where $Z_1\equiv1+\left(1-\frac{a^2}{M^2}\right)^{1/3}\left[\left(1+\frac{a}{M}\right)^{1/3}+\left(1-\frac{a}{M}\right)^{1/3}\right]$ and $Z_{2} \equiv \sqrt{\frac{3a^2}{M^2}+Z_1^2}$.

The ISCO is located at $r_{\rm{ms}}$. We suppose that below ISCO the geodesics are freely falling onto the horizon of the black hole with the energy and angular momentum corresponding to the marginally stable orbit, i.e., the particles keep $E=E_{\rm Kep}(r_{\rm ms})$ and $L=L_{\rm Kep}(r_{\rm ms})$ during their inspiral.

However, neutral elements may undergo a sudden charging process (due to photoionization) at a given radius and obtain nonzero specific charge $q$. While the change of the rest mass $m$ may be neglected, the particle dynamics changes immediately owing to the presence of the magnetic field (\ref{waldpot1}). Values of integrals of motion are changed as follows:
\begin{equation}
\label{newconst}
E=E_{\rm Kep}-qA_t,\;\;\;\;L=L_{\rm Kep}+qA_\varphi.
\end{equation}

\begin{figure*}[htb!]
\centering
\includegraphics[scale=.45]{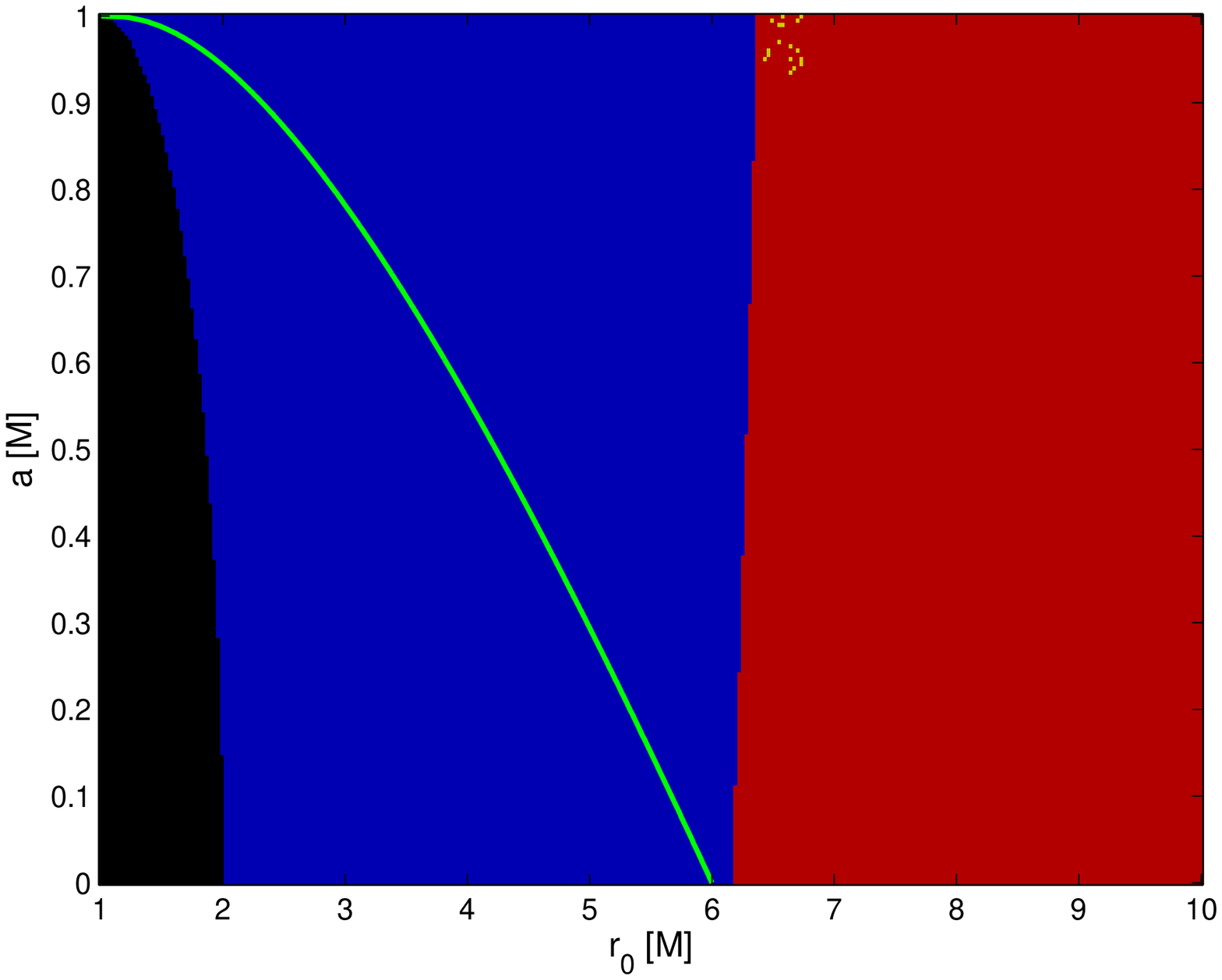}
 \includegraphics[scale=.45]{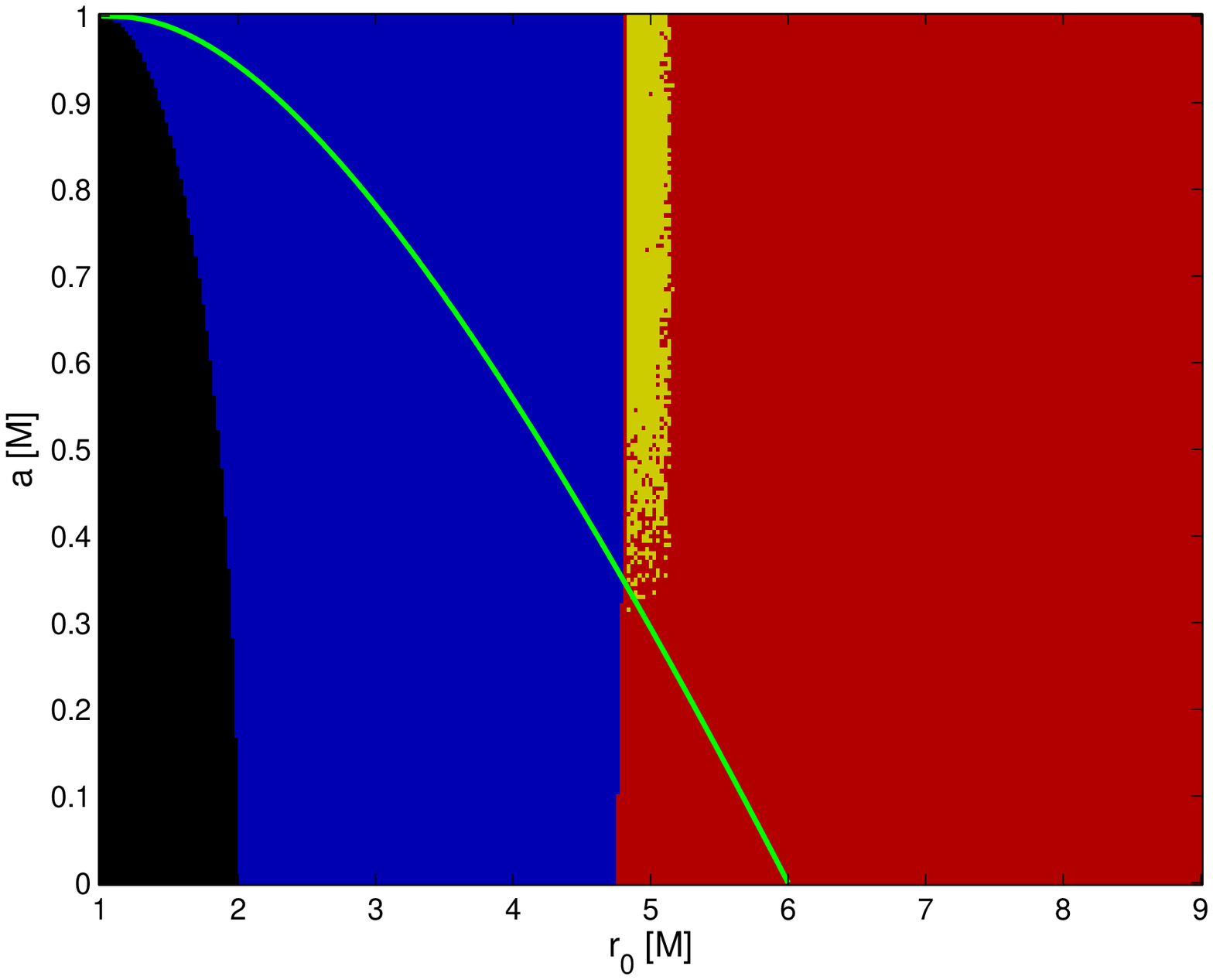}\\
\includegraphics[scale=.45]{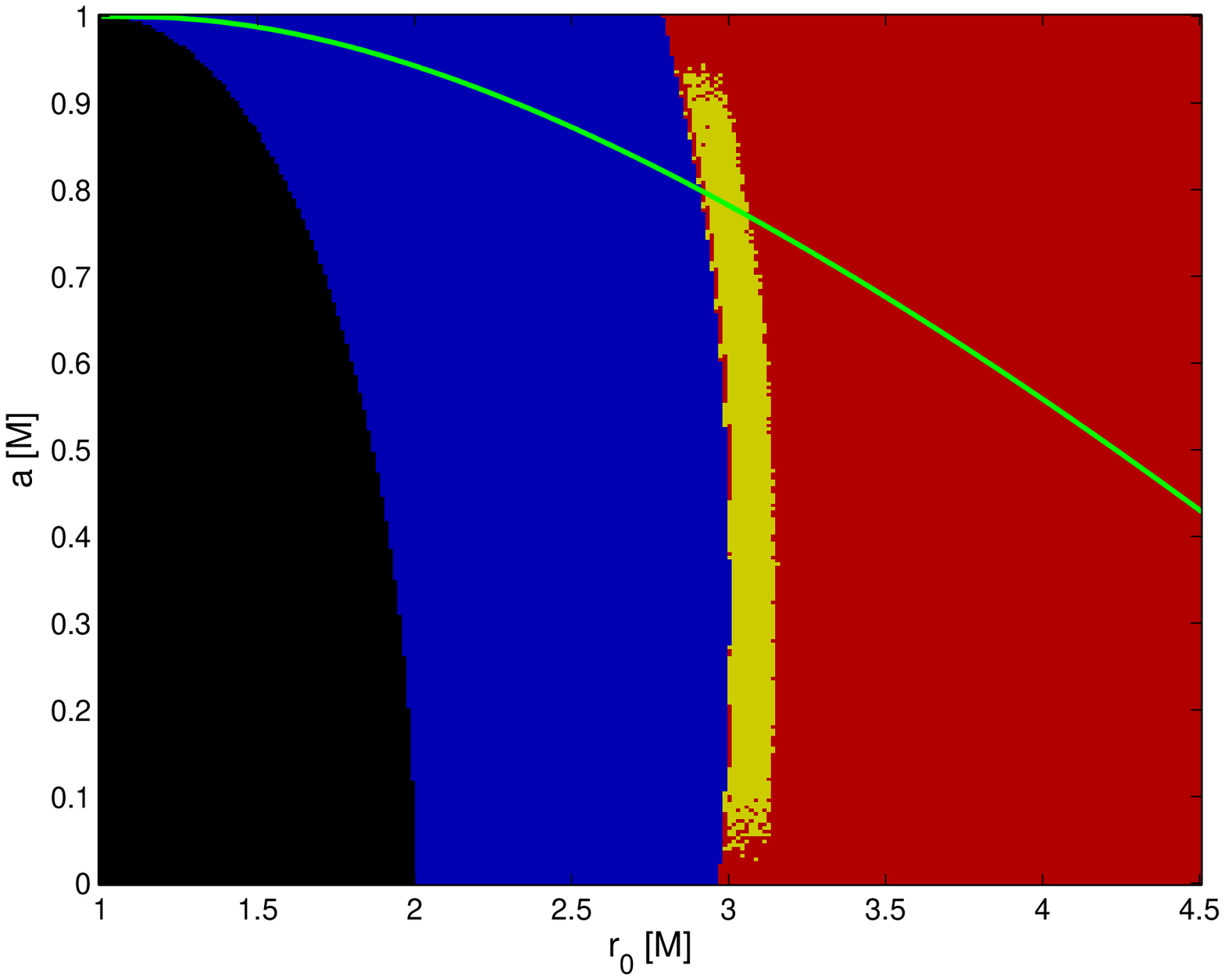}
 \includegraphics[scale=.45]{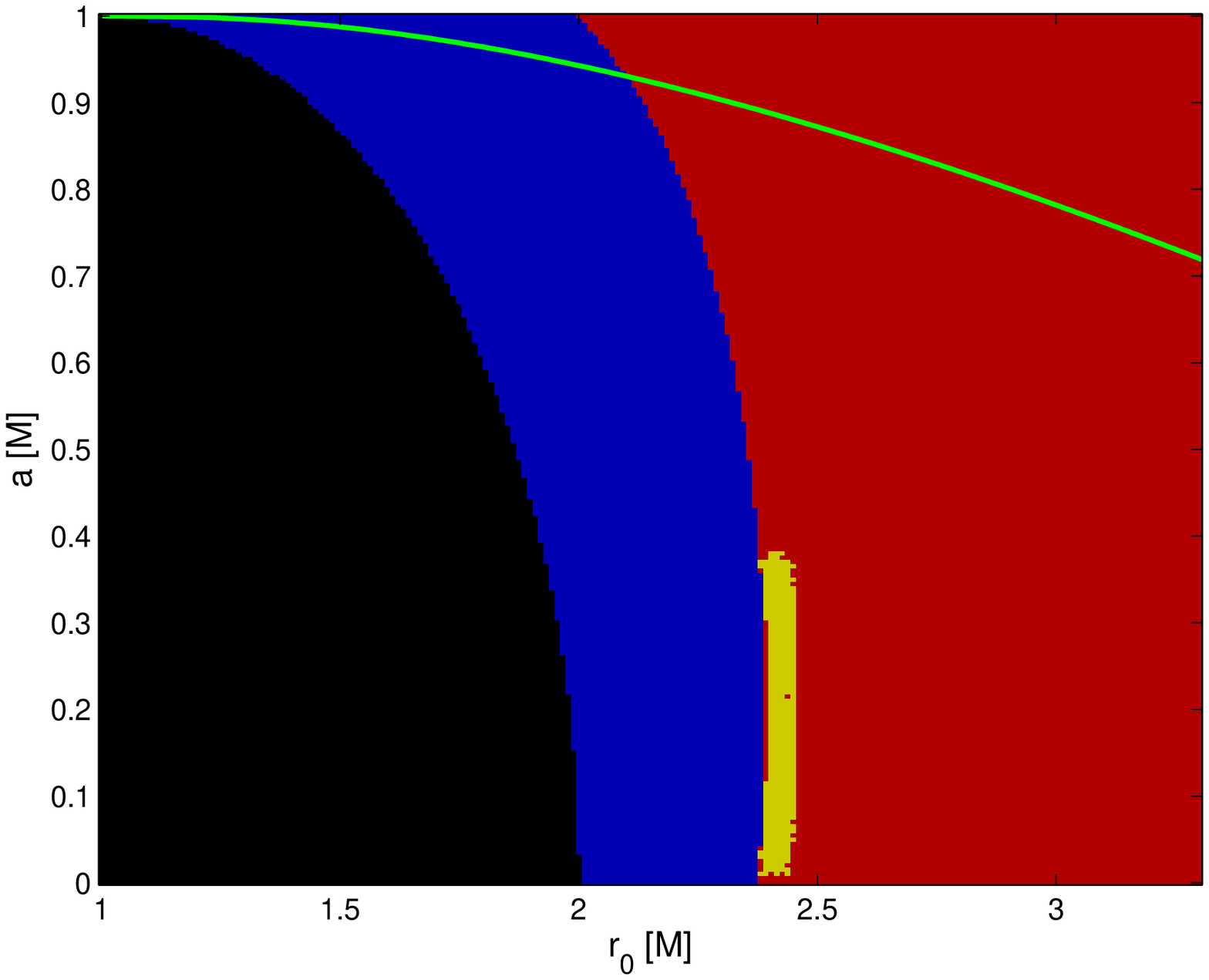}
\caption{Behavior of the particles starting initially at stable circular geodesics (and freely infalling under the ISCO) in the Kerr spacetime after being charged to $qB=-0.5$ (top left), $-1$ (top right), $-5$ (bottom left), and $-20$ (bottom right). Depending on the spin $a$ and the initial radius $r_0$, the particle may fall directly onto the horizon (blue), remain oscillating in the equatorial plane (red), or escape to infinity along the symmetry axis (yellow). Black color denotes the interior of the black hole (region below the outer horizon). The position of the ISCO is indicated by the green line.}
\label{escape_full}
\end{figure*}

The effective potential of the particle changes accordingly. From Eq.~(\ref{eff}) we obtain
\begin{equation}
\label{effdif}
V_{\rm eff}(q)=V_{\rm eff}(q=0)-qA_t,
\end{equation}
which ensures that although the charge shifts the energy of the particle, the difference separating the particle from the allowed minimum (defined by the effective potential) remains unchanged. In particular, stable Keplerian orbits found in local minima of $V_{\rm eff}$ remain on the potential curve even after the ionization. However, they are not located in the potential minima anymore; the former circular orbit is disturbed. As a result, it turns into a stable, radially oscillating (elliptic) orbit, or it may become an unstable (plunging) orbit inspiraling onto the horizon of the black hole. Interestingly, it can also be set on an escaping trajectory. To inspect the latter option, we examine the asymptotic behavior of the effective potential Eq.~(\ref{eff}). For large $r$ we find the dominant term to be given as

\begin{equation}
V_{\rm eff}\propto \frac{|qB|r\sin\theta}{2},
\label{asymeff}
\end{equation}
which shows that particles may only escape to infinity along the axis of symmetry, in which case $r\sin\theta$ remains finite.

In particular, for a particle charged on a Keplerian orbit of radius $r_0$ with former energy $E_{\rm Kep}$ escaping along the symmetry axis, we obtain the following relation valid in the asymptotic region:

\begin{equation}
E-V_{\rm eff}|_{r\gg M}= E_{\rm Kep}-\frac{aqB}{r_0}-\sqrt{1+\frac{q^2B^2r^2\sin{\theta}^2}{4}}.
\label{asym}
\end{equation}

The motion is allowed only for non-negative values of $E-V_{\rm eff}$, and since $E_{\rm Kep}<1$ for finite $r_0$ and $a \geq 0$ is supposed, we observe that  (i) particles may only escape for $qB<0$, (ii) escape is possible only for $a\neq 0$, and (iii) the asymptotic velocity of escaping particles is an increasing function of parameters $|qB|$ and $a$, and a decreasing function of $r_0$.

The first condition, $qB<0$, requires that the charge of the particle is negative for the parallel orientation of the spin and magnetic field and positive for the antiparallel orientation. The escaping orbits are not allowed in the Schwarzschild limit, $a=0$, regardless of the values of $q$, $B$, and $r_0$. The asymptotic velocity of escaping particles rises with increasing values of $a$, $|q|$, and $|B|$ and will be higher for the particles charged closer to the horizon.

Thus, we conclude that escaping orbits are in principle possible, i.e., that necessary conditions for the escape may be fulfilled  in a given setup. However, these conditions are not sufficient, and they give no guarantee that trajectories escaping from the equatorial plane are actually realized. To test whether ionized particles really escape, and for the discussion of their asymptotic velocity, we  need to switch to the numerical approach and integrate particular trajectories.

\begin{figure*}[ht]
\centering
 \includegraphics[scale=.45]{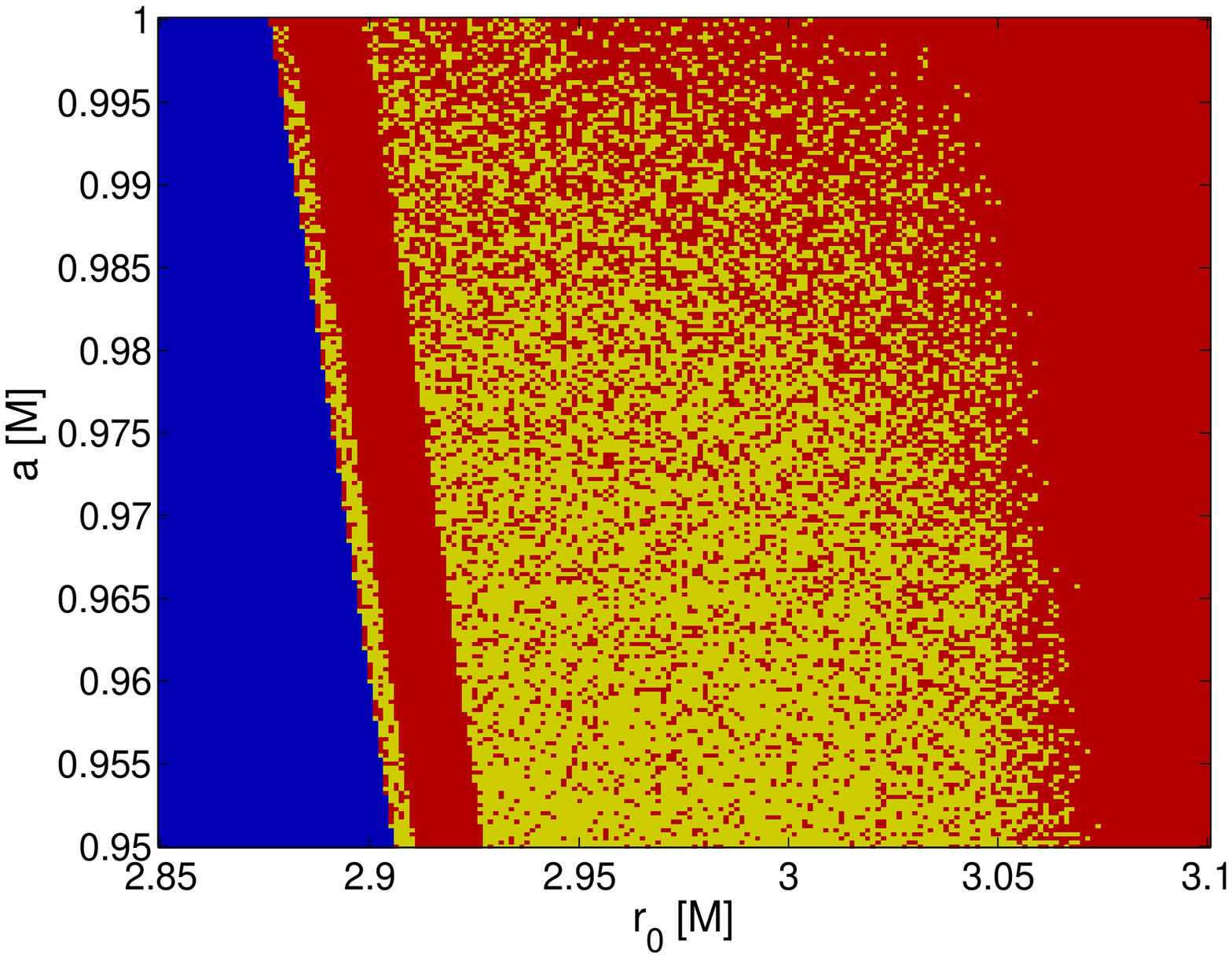}
 \includegraphics[scale=.45]{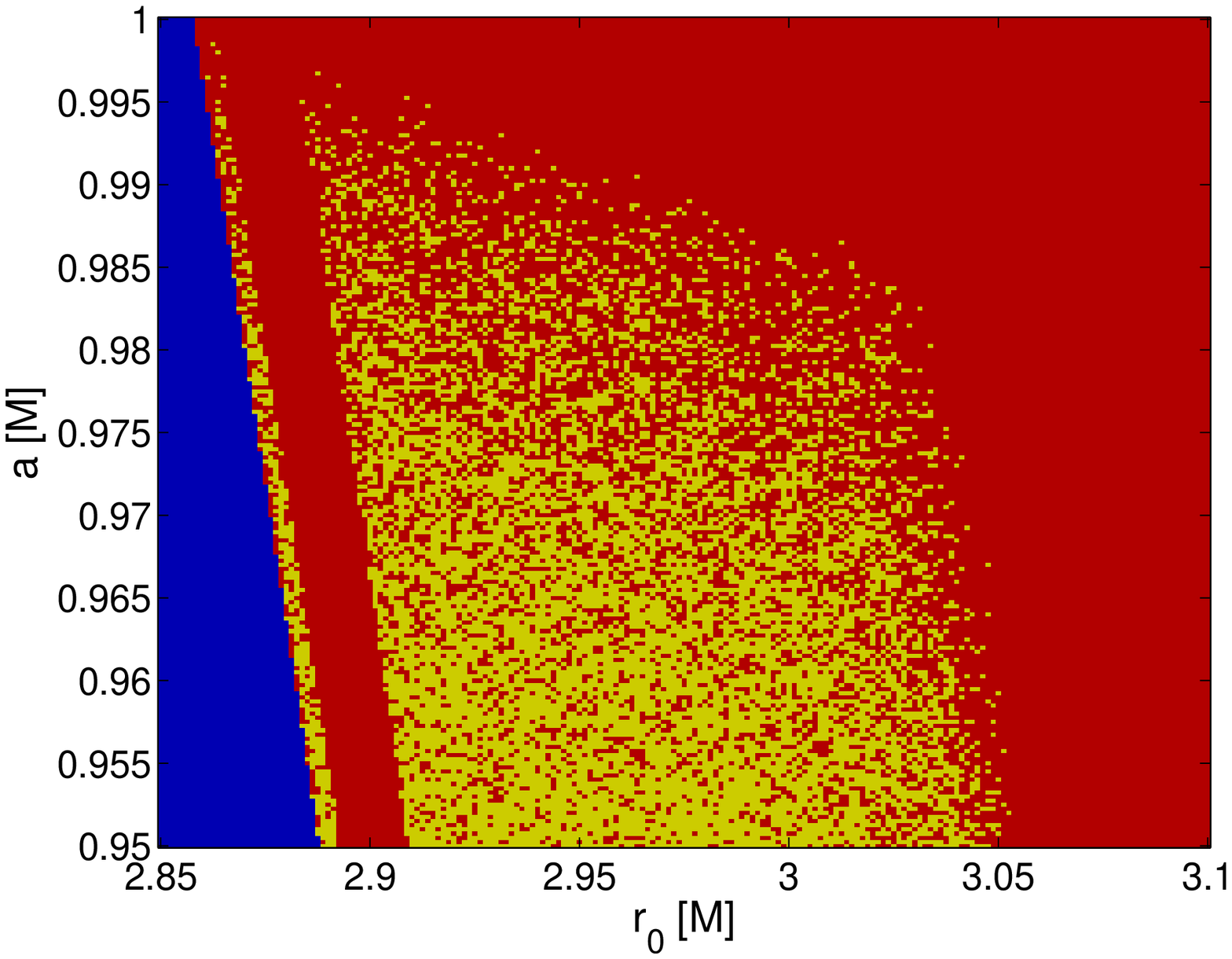}
\caption{High-spin edge of the escape zone for $qB=-4.5$ (left panel) and $qB=-4.6$ (right panel), respectively, is shown here to demonstrate that the escape zone disconnects from the extremal spin value. Color coding is the same as in the previous figure. Let us note that the region of escaping trajectories is not continuous, as is also seen in this plot. The parameter values in the $r_0\times a$ plane (radius vs. spin) for escaping trajectories (denoted by yellow color) are intermixed with oscillating captured trajectories (red) at every level of resolution of the plot.}
\label{escape_odpojeni}
\end{figure*}

\begin{figure*}
\centering
\includegraphics[scale=.45]{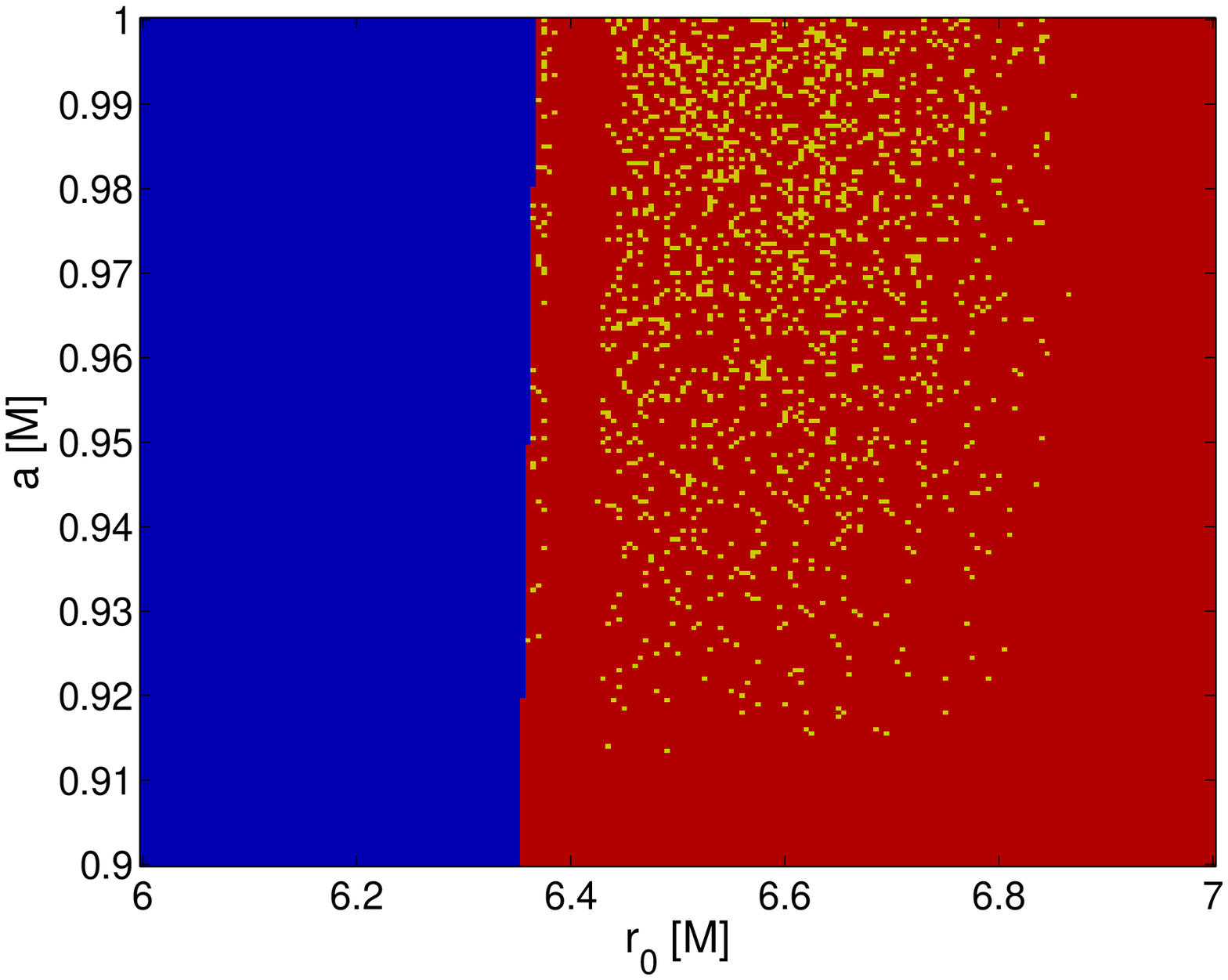}
 \includegraphics[scale=.45]{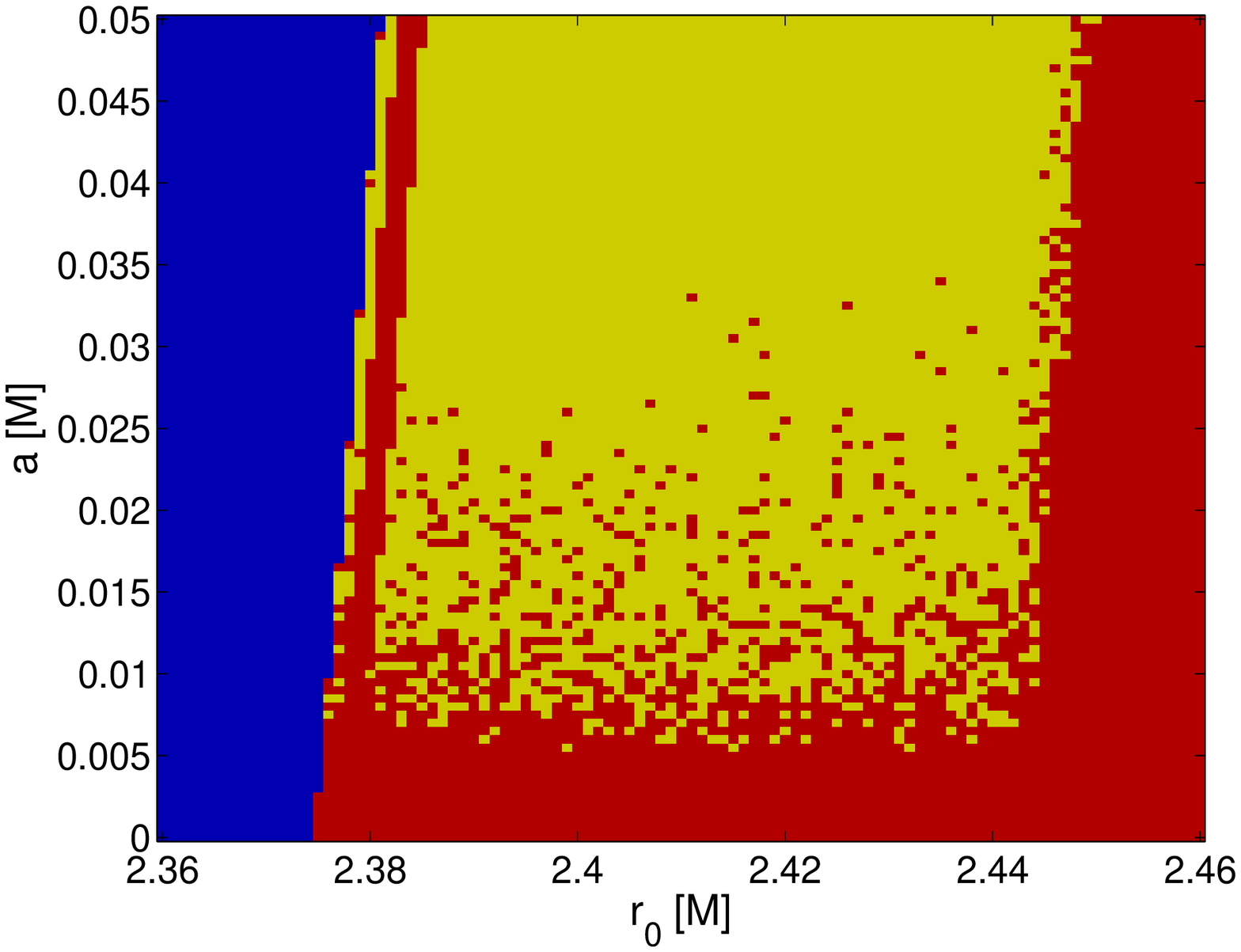}
\caption{Left panel: detail of escape zone for $qB=-0.5$. Right panel: exploration of the low-spin end of the escape zone for $qB=-20$. Color coding is the same as in Fig.~\ref{escape_full}.}
\label{escape_tail}
\end{figure*}

\subsection{Formation of the Escape Zone}
\label{escape_zone}
The crucial parameters that determine the behavior of the Keplerian particle after it obtains electric charge are the black hole spin $a$, the ionization radius $r_0$, and the product $qB$ (hereafter, {\em magnetization parameter})\footnote{The electric charge of the particle and the magnetic field strength do not appear independently in the equations of motion, as we only consider weak magnetic fields in the test-field approximation, i.e.,  the field's effect on the spacetime geometry is neglected and $B$ does not appear in the metric tensor $g_{\mu\nu}$.}. In the rest of the paper we scale all quantities by the black hole's rest mass $M$, which corresponds to setting $M=1$ in the equations and leaves all the quantities dimensionless.

In the search of escaping particles we need to cover an adequate range of values of crucial parameters. Therefore, we set up a grid where the vertical axis will correspond to the spin in the range $a\in\left<0,1\right>$ and the horizontal axis will represent the ionization radius $r_0\in\left<1,r_{\rm out}\right>$, where $r_{\rm out}$ is chosen such that we do not omit any potentially interesting part of the accretion disk. The number of data points of the grid would be typically $200 \times 200$, and we integrate the trajectory corresponding to each point of the grid with a fixed value of $qB=const$. Each integration is set to run up to the value of the affine parameter $\lambda=2\times10^3$, which corresponds to $\approx 100$ azimuthal periods of the original Keplerian orbit near the ISCO. The threshold above which the particle is considered escaping is set to $r=10^2$. On the other hand, the orbit is regarded as plunging if it closely approaches the horizon $r=r_{+}+10^{-4}$. If the final radius lies between the both limits, the respective particle is considered stable. 

\begin{figure*}[ht]
\centering
\includegraphics[scale=.45]{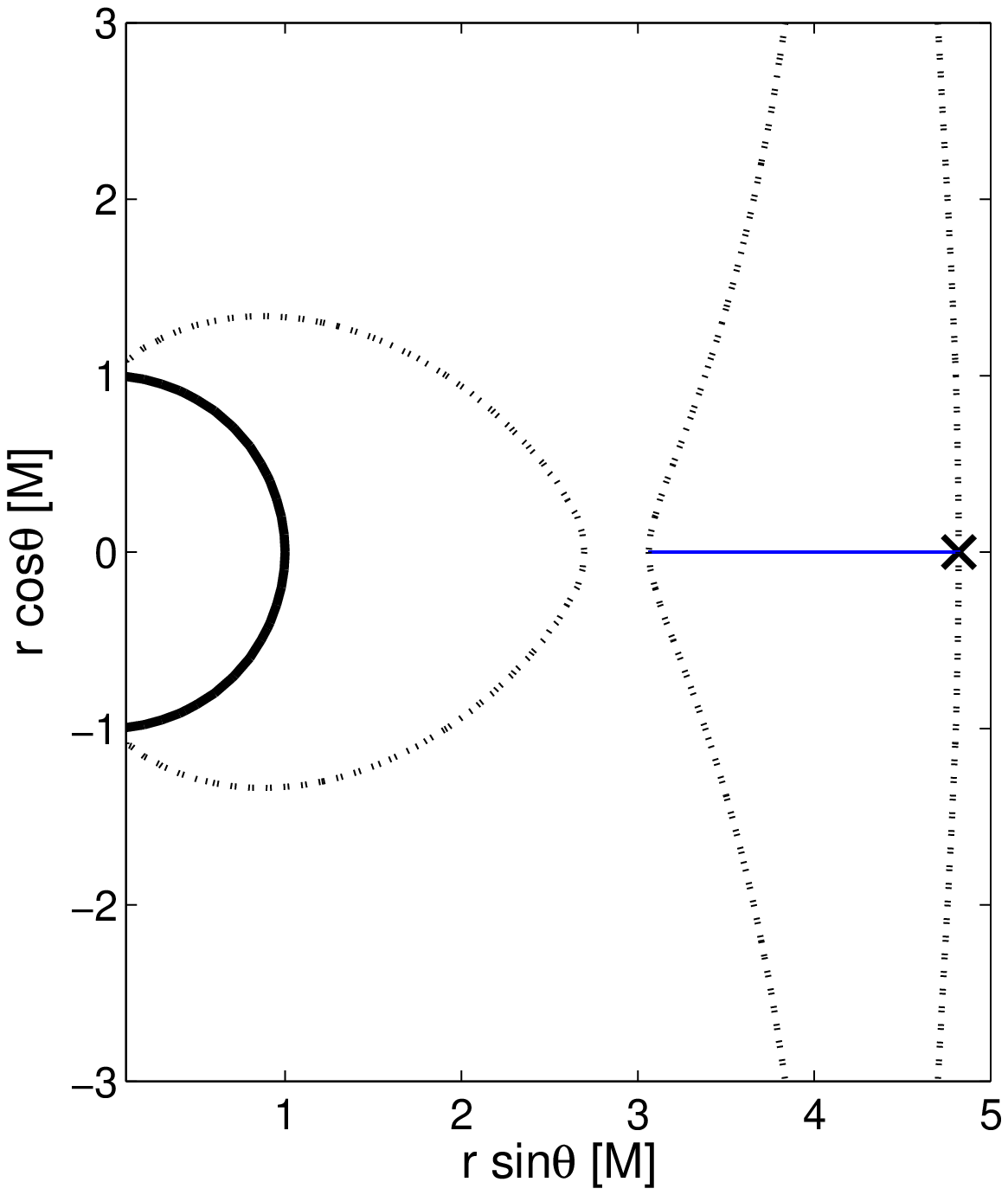}
\includegraphics[scale=.45]{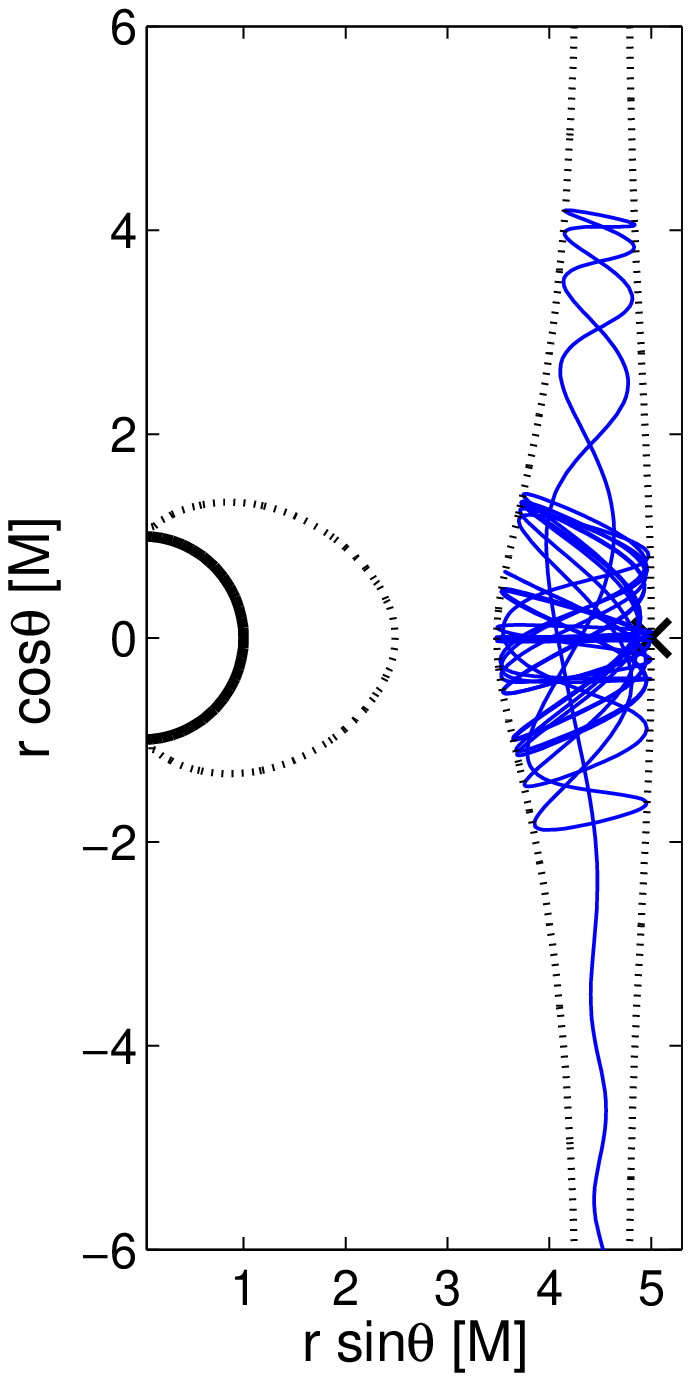}
\includegraphics[scale=.45]{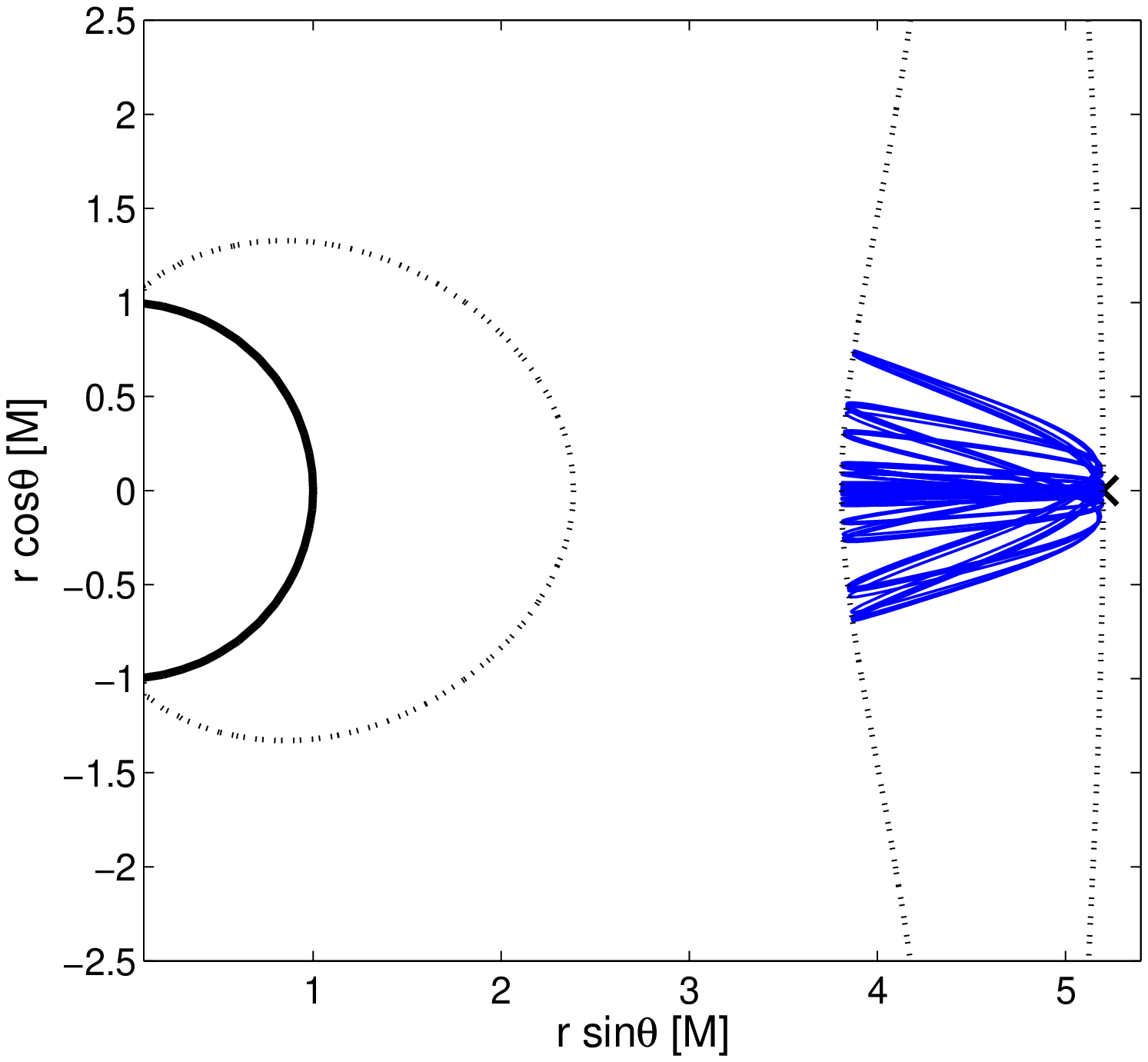}
\caption{Typical trajectories encountered in the escape zone. The left panel ($r_0=4.82$) shows the stable orbit oscillating in the equatorial plane, the middle panel illustrates the escaping particle charged at $r_0=5$, while the right panel shows the stable orbit oscillating around the equatorial plane ($r_0=5.2$). Parameters of the system are $a=1$ and $qB=-1$. The initial radius of the Keplerian orbit is marked by the cross. The solid black line denotes the horizon of the black hole. Dashed lines show relevant isopotential contours of the effective potential. The allowed region for escaping trajectories is bound between the two rotational iso-surfaces forming the open corridor along the symmetry axis ($z\equiv r\cos\theta$). In the equatorial plane the inner iso-surface bends toward the horizon, but they both become perfectly cylindrical and parallel farther from the black hole}
\label{traj_pot}
\end{figure*}

\begin{figure*}[ht]
\centering
\includegraphics[scale=.38]{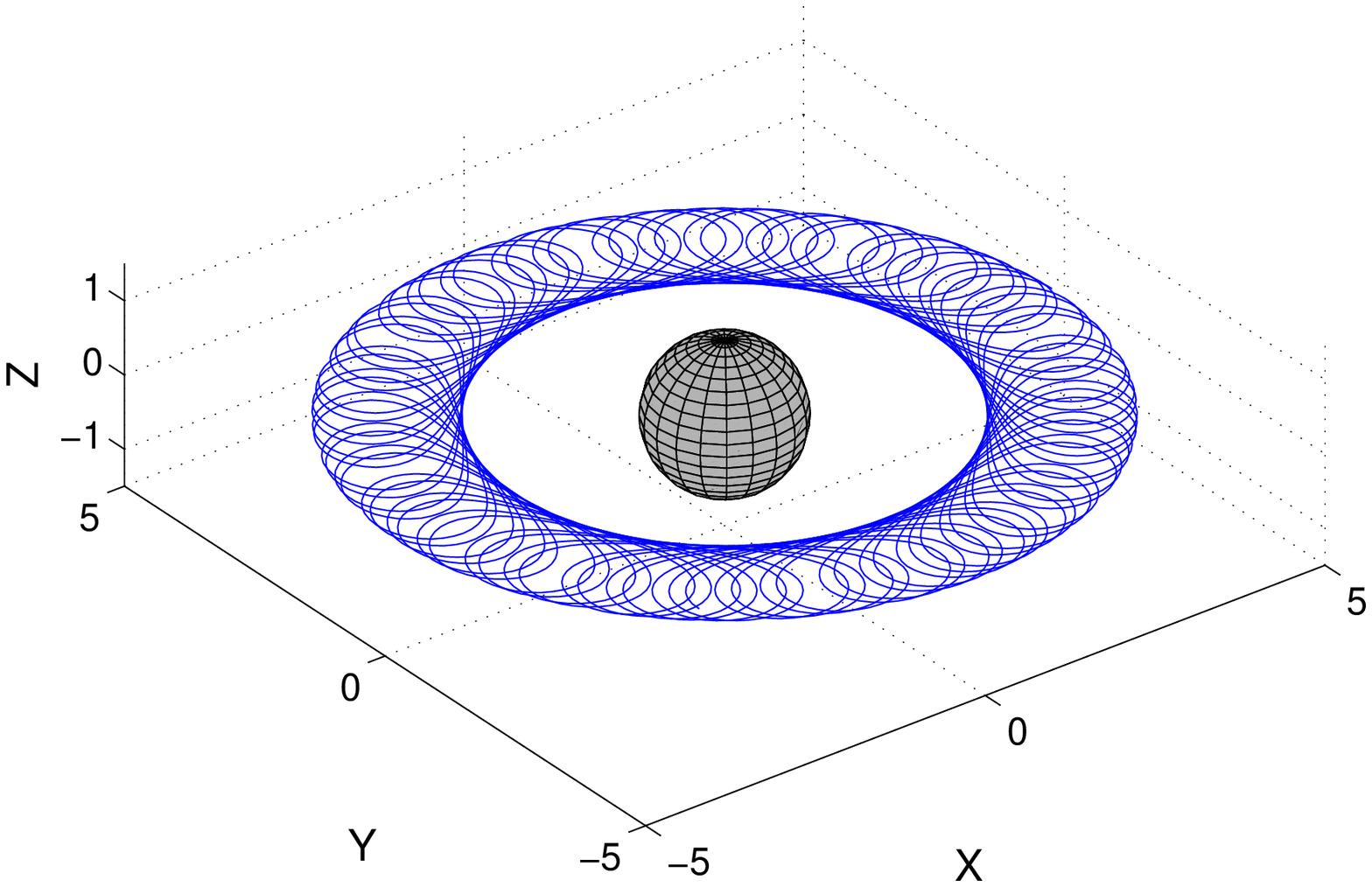}
\includegraphics[scale=.305]{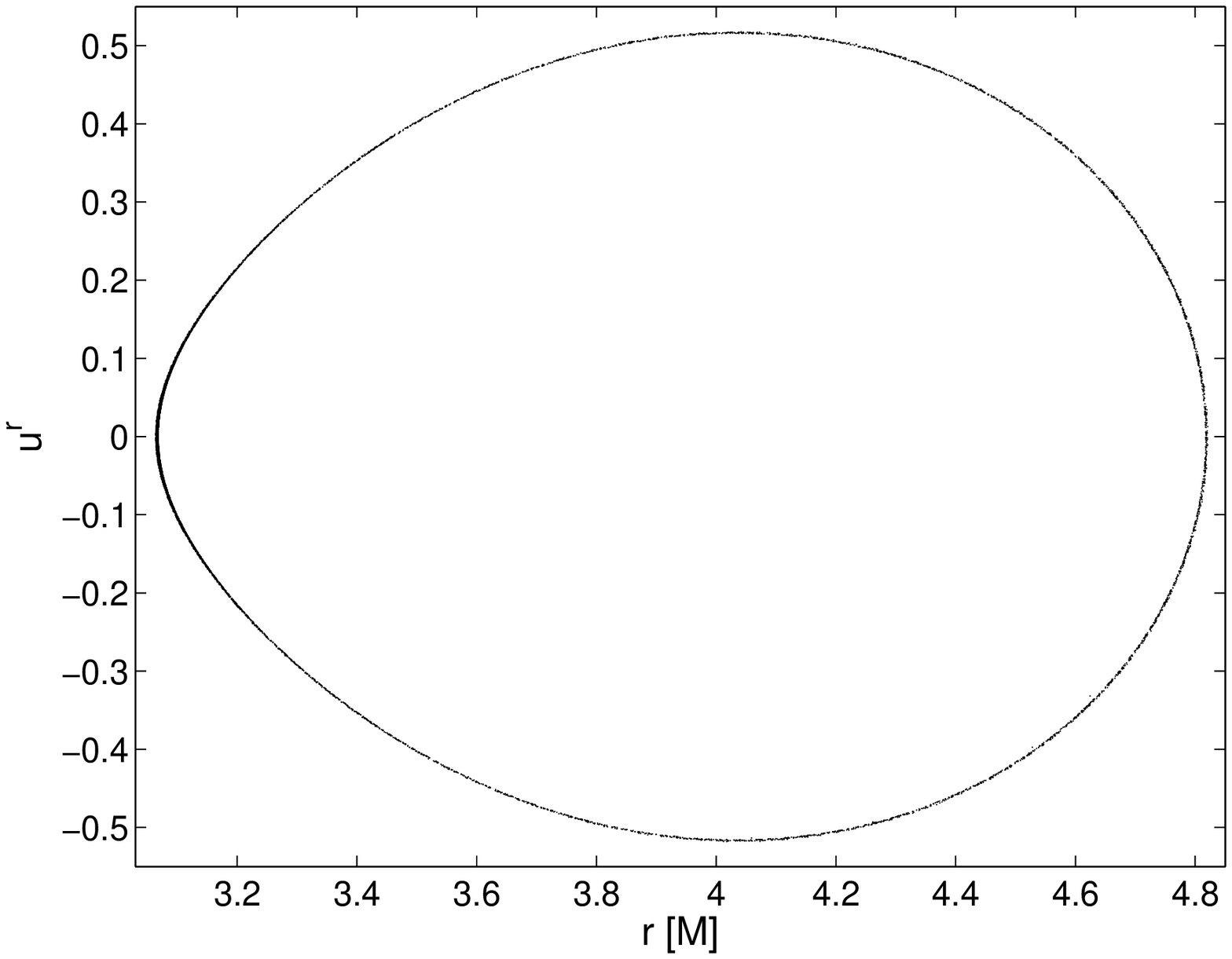}
\includegraphics[scale=.322]{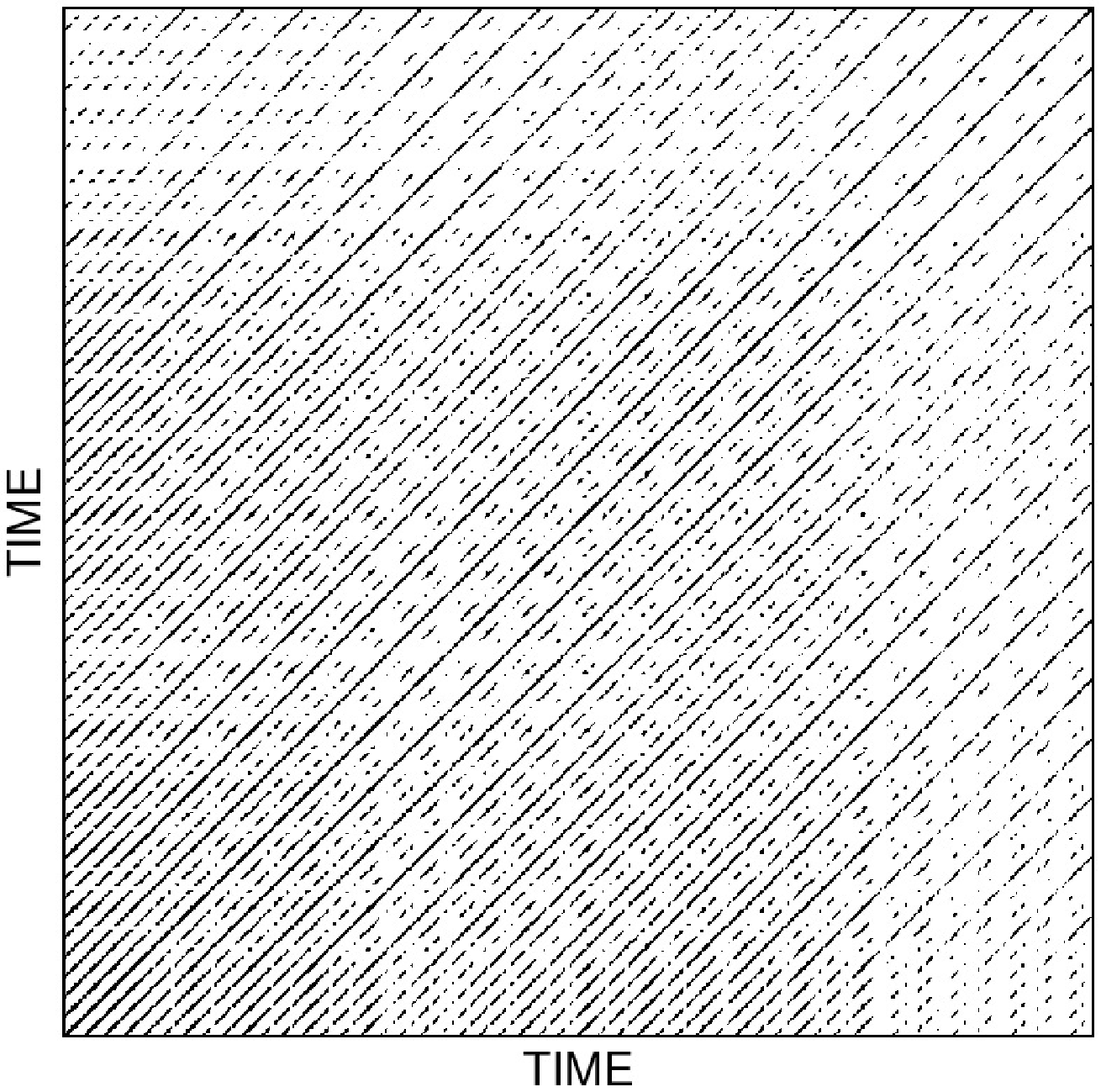}
\caption{An example of a stable orbit (with $r_0=4.82, a=1$ and $qB=-1$; same as in the left panel of Fig.~\ref{traj_pot}) is shown (left panel; gray surface marks the horizon), and its Poincar\'{e} section is constructed (middle panel). In the right panel we show the recurrence plot of this trajectory ($\varepsilon=1.1$).}
\label{traj_stable}
\end{figure*}

Results of the integration are visualized by assigning different colors to grid points corresponding to trajectories with different final states. This type of graphic representation of the dynamic system  is usually denoted as a basin-boundary plot \citep[see, e.g.,][where basin-boundary plots are applied in a similar context]{frolov10,alzahrani13,alzahrani14}. We use blue color for plunging orbits, red for stable orbits, and yellow for escaping ones. We begin with the geodesic limit $qB=0$, for which no escapes are allowed and the borderline between the blue plunging region and the red area of stable orbits is given by the ISCO curve. The analysis in Sec.~\ref{ionization} shows that only particles with $qB<0$ may acquire sufficient energy to reach infinity. Thus, we introduce the charge and set $qB=-0.1$. The borderline between the regions is no longer defined by the ISCO. In general, the charge may cause the originally stable orbit to become plunging and vice versa, as some plunging neutral particles become stabilized by the ionization during their infall below the ISCO. Nevertheless, no escapes are observed for $qB=-0.1$.

We gradually decrease the value of $qB$ and eventually detect escaping orbits for $qB=-0.5$ (top left panel of Fig.~\ref{escape_full}). For this case, the escaping orbits are only allowed for spin values close to the extremal value $a=1$, and they spread rather sporadically near the borderline separating the blue infall zone from the red area of stable orbits. In the top right panel of Fig.~\ref{escape_full} we present the plot for $qB=-1$, for which the density of escaping trajectories substantially increases and they form a narrow escape zone that extends down to spin values as low as $a\approx0.3$. This plot also illustrates that even freely falling particles below the ISCO may still escape the attraction of the center owing to the charging process. Switching to $qB=-5$ (bottom left panel of Fig.~\ref{escape_full}) results in shifting the escape zone further to lower spins and simultaneously to lower radial values, i.e., closer to the horizon. We also observe that for this value of $qB$ the upper end of the escape zone detaches from the $a=1$ limit and escape of particles from the vicinity of the black hole with spin close to extremal is not permitted anymore. Increasing the magnitude of the magnetization parameter to $qB=-20$ (bottom right panel of Fig.~\ref{escape_full}) confirms both these trends and the escape zone shifts further down to lower spin (though it can never reach the static limit $a=0$) and becomes generally closer to the horizon.
 
Properties of the escape zone are further explored in Figs.~\ref{escape_odpojeni} and \ref{escape_tail}.  In the former we observe in detail that the escape zone detaches from  the extremal spin $a=1$ as $qB$ decreases. Indeed, while for $qB=-4.5$ there are still some escaping particles with $a=1$ (left panel of Fig.~\ref{escape_odpojeni}), none of them remain if $qB$ decreases to $qB=-4.6$ (right panel of Fig.~\ref{escape_odpojeni}). We conclude that for $qB\lessapprox -4.5$ the escaping trajectories are only realized for $a<1$. In Fig.~\ref{escape_tail} we show the effect of the magnetization parameter and spin on escaping orbits from yet another perspective. While for small values of $|qB|$ the escape is only possible for spin close to the extremal value (left panel of Fig.~\ref{escape_tail} with $qB=-0.5$), for higher $|qB|$ the escape zone is shifted toward lower values of spin, escape is allowed  even for very low $a$, and high spins are actually excluded (right panel of Fig.~\ref{escape_tail} with $qB=-20$).

\begin{figure*}[ht]
\centering
\includegraphics[scale=.52]{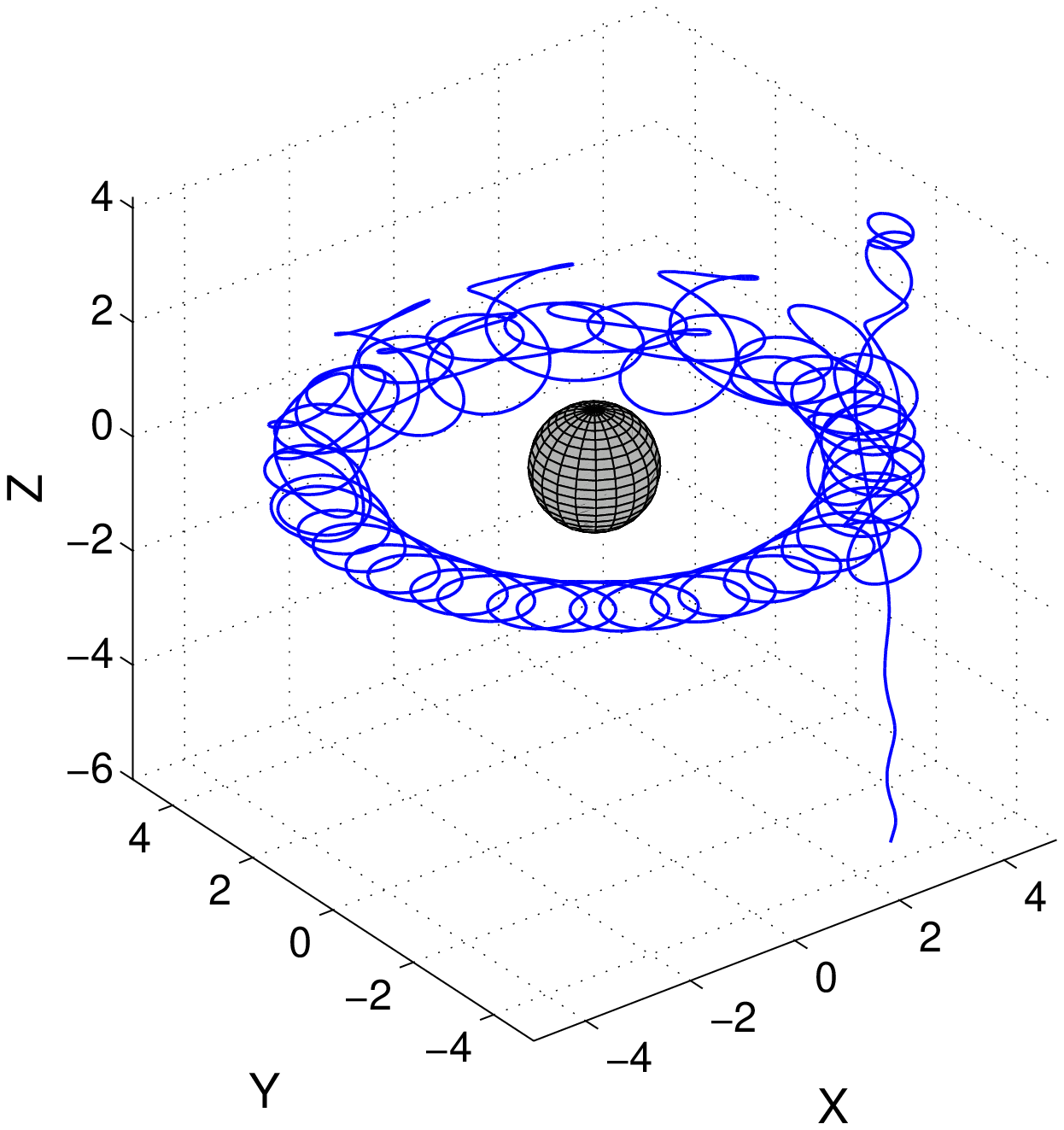}
\includegraphics[scale=.5]{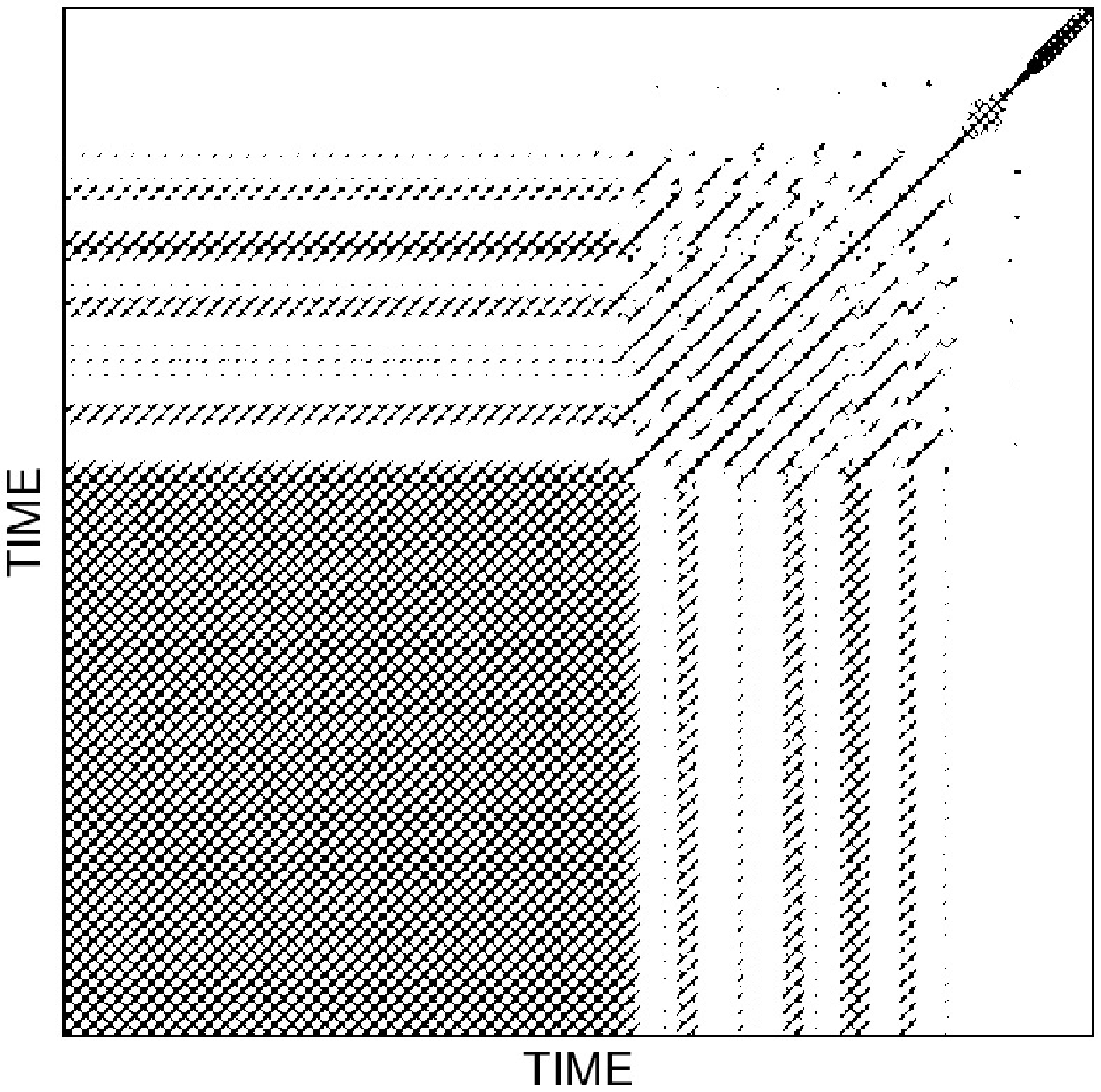}
\caption{An escaping orbit (with $r_0=5, a=1$ and $qB=-1$; same as in the middle panel of Fig.~\ref{traj_pot}) is shown in a 3D projection (left panel; gray surface marks the horizon). In the right panel we show the recurrence plot of this trajectory.}
\label{traj_escape}
\end{figure*}

In order to illustrate the dynamics of stable and escaping orbits in the escape zone, we plot the poloidal projection of typical trajectories along with the corresponding isopotential contours in Fig.~\ref{traj_pot}. In accordance with the results of the analysis in Sec.~\ref{ionization}, we observe that the isopotentials for the negatively charged particles do not close and create the infinite narrow corridor parallel to the rotation axis, making the escape in principle possible. However, only some particles do actually escape (such as the one in the middle panel of Fig.~\ref{traj_pot}). In the left and the right panels we observe that the particle may also remain oscillating on the stable orbit in or around the equatorial plane, depending on the initial radius $r_0$. To inspect the dynamics of stable orbits we can employ standard techniques -- e.g., Lyapunov characteristic exponents \citep{skokos10} or Poincar\'{e} sections. The latter is applied in Fig.~\ref{traj_stable} (middle panel) for the case of orbit with $r_0=4.82, a=1$, and  $qB=-1$ (the same orbit as in the left panel of Fig.~\ref{traj_pot}).  The trajectory is integrated up to $\lambda_{\rm fin}=10^5$, resulting in $\approx 10^4$ section points that form a closed curve typical for a regular quasi-periodic orbit. Nevertheless, in the case of escaping trajectory  these standard tools are not applicable, as the particle escapes too early, leaving only a few section points in the equatorial plane, which do not allow us to decide whether they form a curve or rather a surface. Similarly, the computation of the Lyapunov characteristic exponents requires a much longer integration period to converge (denoted as {\em Lyapunov time}, \citet{skokos10}). However, we can still apply recurrence plots \citep{marwan07}, which are known to reflect the characteristic dynamic properties on a much shorter time scale. Recurrence plots (RPs) represent the recurrences of the trajectory to the defined $\varepsilon$ neighborhood in the phase space. Visual survey of the RPs allows us to discriminate between  the dynamic regimes of the system. In particular, regular orbits generally form diagonal patterns in the RP as shown for the stable orbit in the right panel of Fig.~\ref{traj_stable}. On the other hand, chaotic dynamics is characterized by disrupted diagonal patterns as observed for the escaping trajectory in the right panel of Fig.~\ref{traj_escape}. The given particle is integrated up to $\lambda_{\rm fin}=575$ (when its escape begins), which is sufficient to produce an indicative RP.  For the details on recurrence analysis we refer to the review by \citet{marwan07}, while the full description of the application of RPs in the analogous context may be found in our previous paper \citep{kopacek10}. For a recent application of this method in relativistic astrophysics, see \citet{semerak12,witzany15,sukova16} and \citet{lukes18}.

\begin{figure*}[ht]
\centering
 \includegraphics[scale=.45]{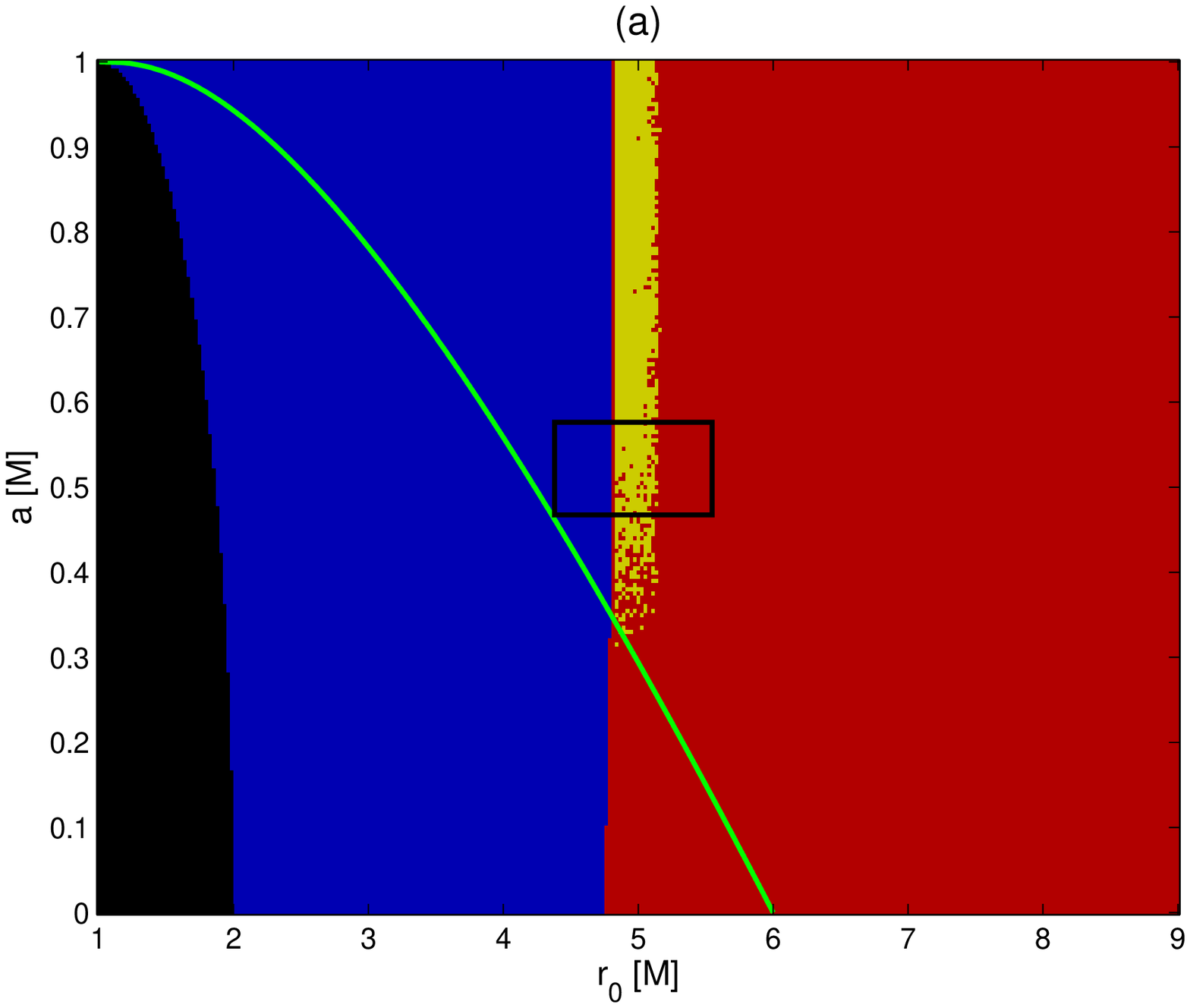}
 \includegraphics[scale=.45]{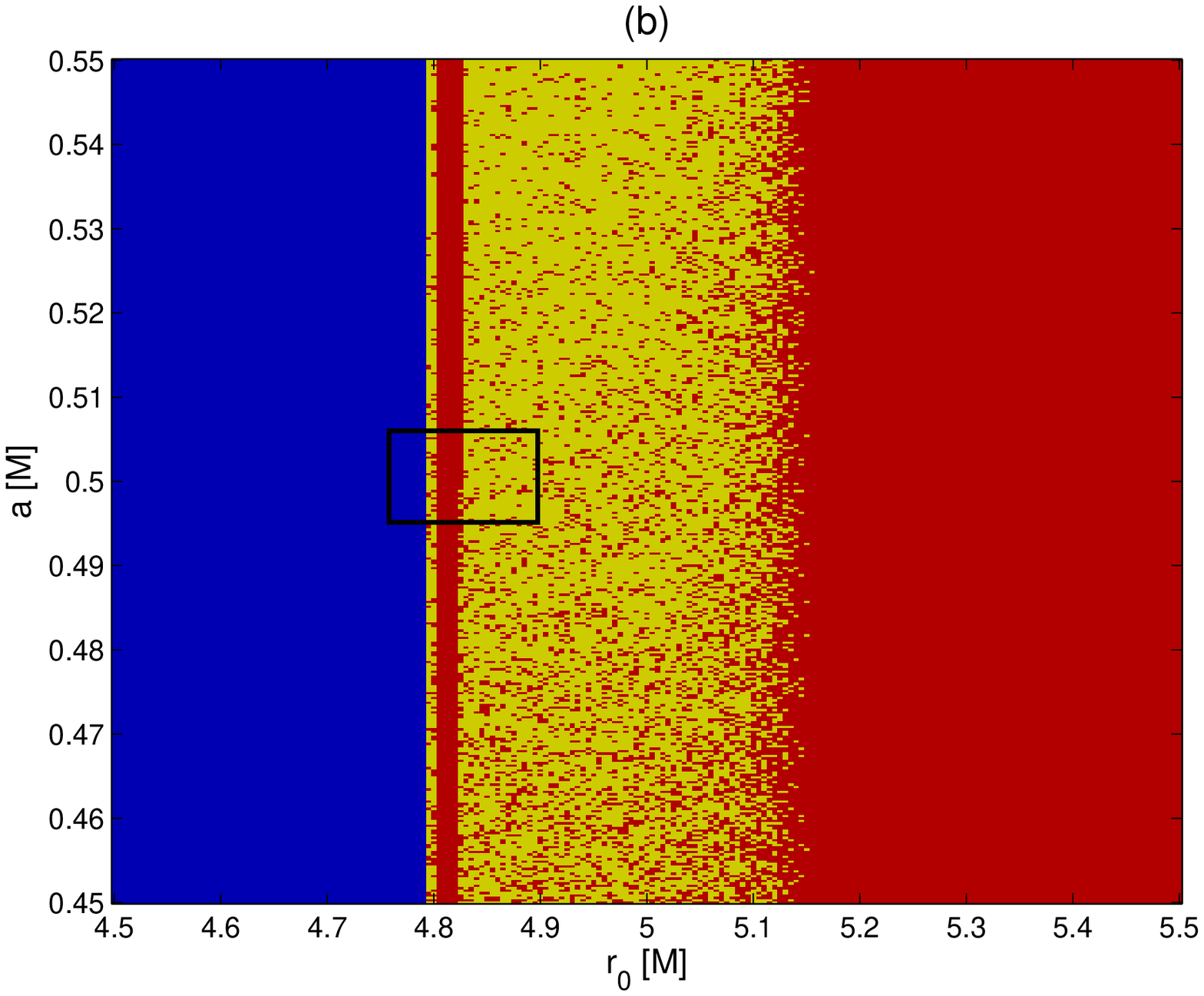}\\
\includegraphics[scale=.45]{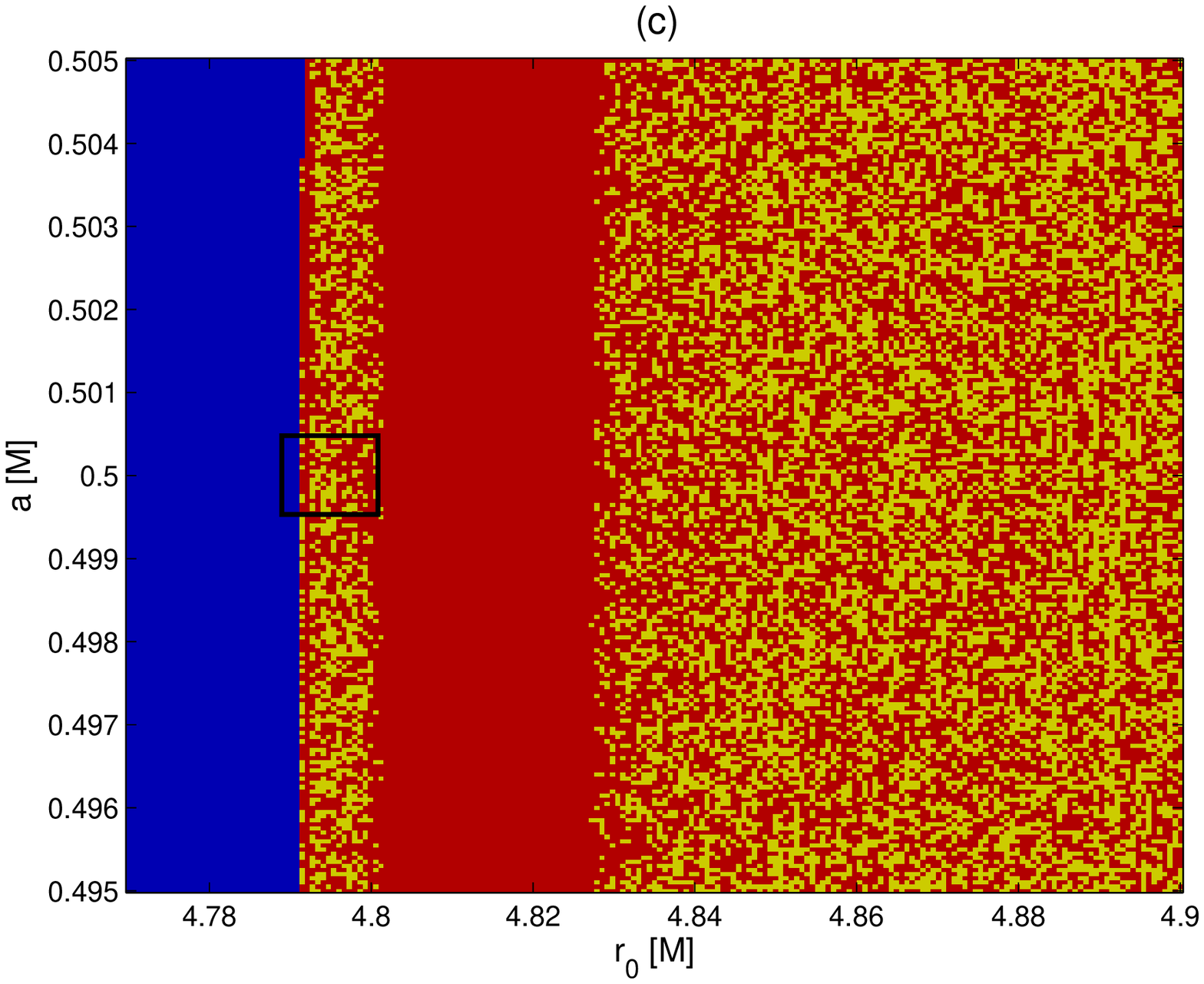}
 ~\includegraphics[scale=.45]{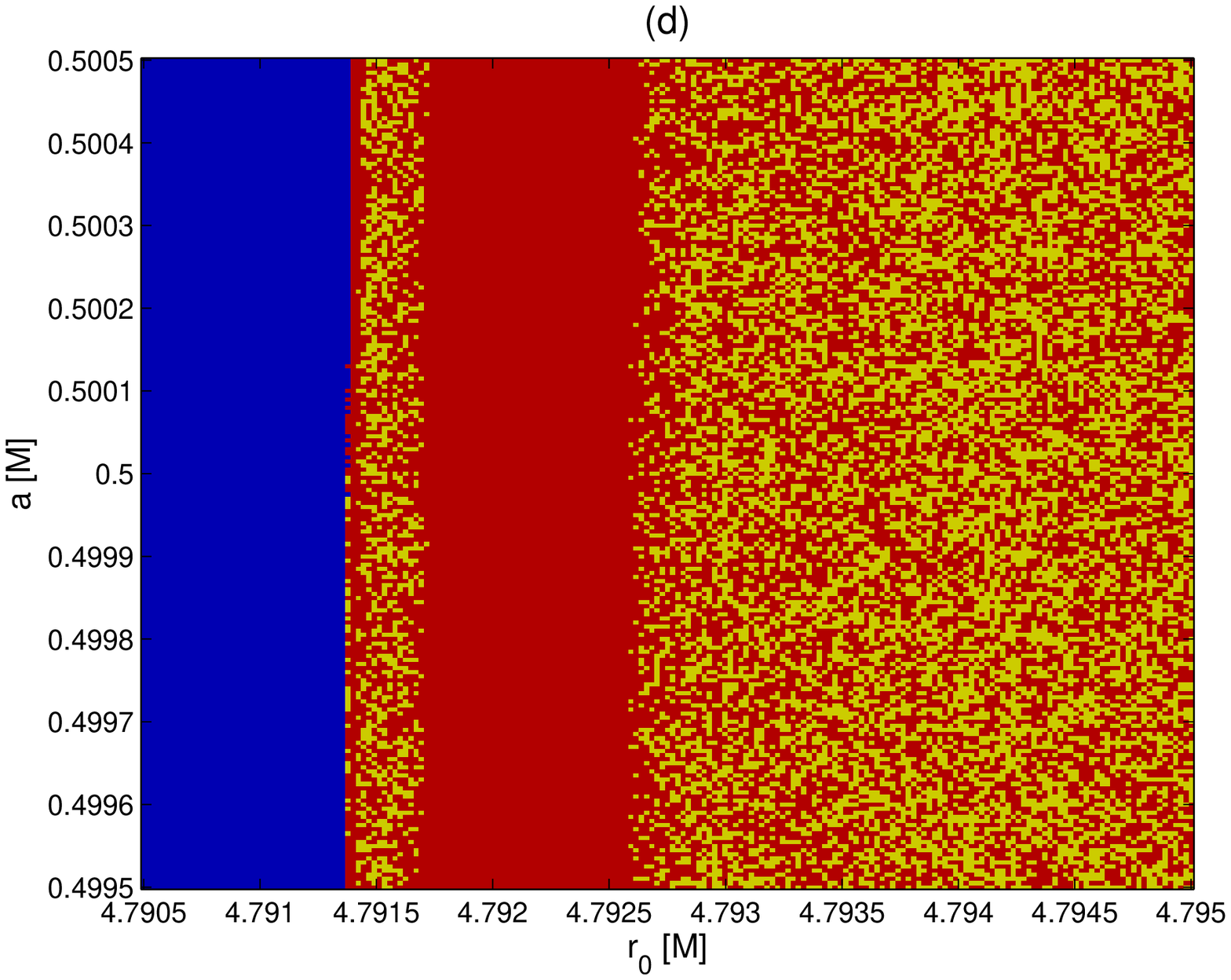}
\caption{The fine structure of the escape zone for the case $qB=-1$ is explored. Going from the top left to the bottom right panel, the portions of the plots marked by the black rectangles are  magnified in the subsequent panels, showing self-similarity in the structure of the escape zone.}
\label{escape_detail2}
\end{figure*}

From the above we conclude that escaping particles fundamentally differ from the stable ones in their dynamic regime. In particular, stable orbits are regular (quasi-periodic), while the escaping particles show hallmarks of the deterministic chaos. Chaotic orbits generally tend to ergodically fill the allowed region, which, in our case,  gets them sooner or later  on the escaping trajectory. The defining property of the deterministic chaos is the sensitivity to the initial condition,  which the escaping particles exhibit in a particularly spectacular manner -- we observe that the orientation of the escaping particle (i.e., whether it escapes upward or downward from the equatorial plane) is sensitive to the initial condition, and even the tiniest shift of the parameters may switch the orientation of the outflow.   

We have revealed the dynamic nature of both the stable and escaping trajectories, which allows us to clarify the complex structure of the escape zone. In Fig.~\ref{escape_detail2} we explore the particular escape zone (with $qB=-1$) in detail. The region inside the zone is densely populated by both the chaotic escaping orbits and the regular quasi-periodic orbits that remain oscillating in (or around) the equatorial plane. The density of escaping trajectories generally increases with the spin. We notice the red strip of stability close to the inner edge of the escape corridor, which corresponds to the main resonance of fundamental frequencies of the system \citep[e.g., ][]{contopoulos02}. If we magnify this section of the escape zone, we find minor stability strips corresponding to the higher-order resonances.  Self-similarity is observed in the distribution of the escaping orbits. Such a behavior is typical for the dynamical system with a nonintegrable perturbation (which is introduced by the the electric charge and magnetic field in our case). Dynamics of the perturbed system may be understood using the general results of the chaos theory \citep[e.g.,][]{lichtenberg92,tel06}. In particular, the fundamental Kolmogorov--Arnold--Moser theorem (KAM theorem) and Poincar\'{e}-Birkhoff theorem show why some orbits resist the perturbation while others turn chaotic, and they explain the behavior of the perturbed system close to resonances.

\subsection{Acceleration of Escaping Particles}
\label{acceleration}
In the previous section we explored the formation of the escape zone and discussed its location in the $r_0 \times a$ plane with respect to the magnetization parameter $qB$. Here we investigate the final velocity of escaping particles and, in particular, we seek the most accelerated ones. In order to quantitatively discuss the acceleration of escaping particles of four-velocity $u^{\mu}$, we determine its linear velocity $v^{(i)}$ with respect to the locally nonrotating frame \citep{bardeen72} with the tetrad basis $e^{(i)}_{\mu}$ as follows:
\begin{equation}
\label{linspeed}
 v^{(i)}=\frac{u^{(i)}}{u^{(t)}}=\frac{e^{(i)}_{\mu}u^{\mu}}{e^{(t)}_{\mu}u^{\mu}}
\end{equation}
and we use it to calculate Lorentz factor $\gamma=(1-v^2)^{-1/2}$ where $v=\sqrt{[v^{(r)}]^2+[v^{(\theta)}]^2+[v^{(\varphi)}]^2}$.

\begin{figure*}[ht]
\centering
\includegraphics[scale=.45]{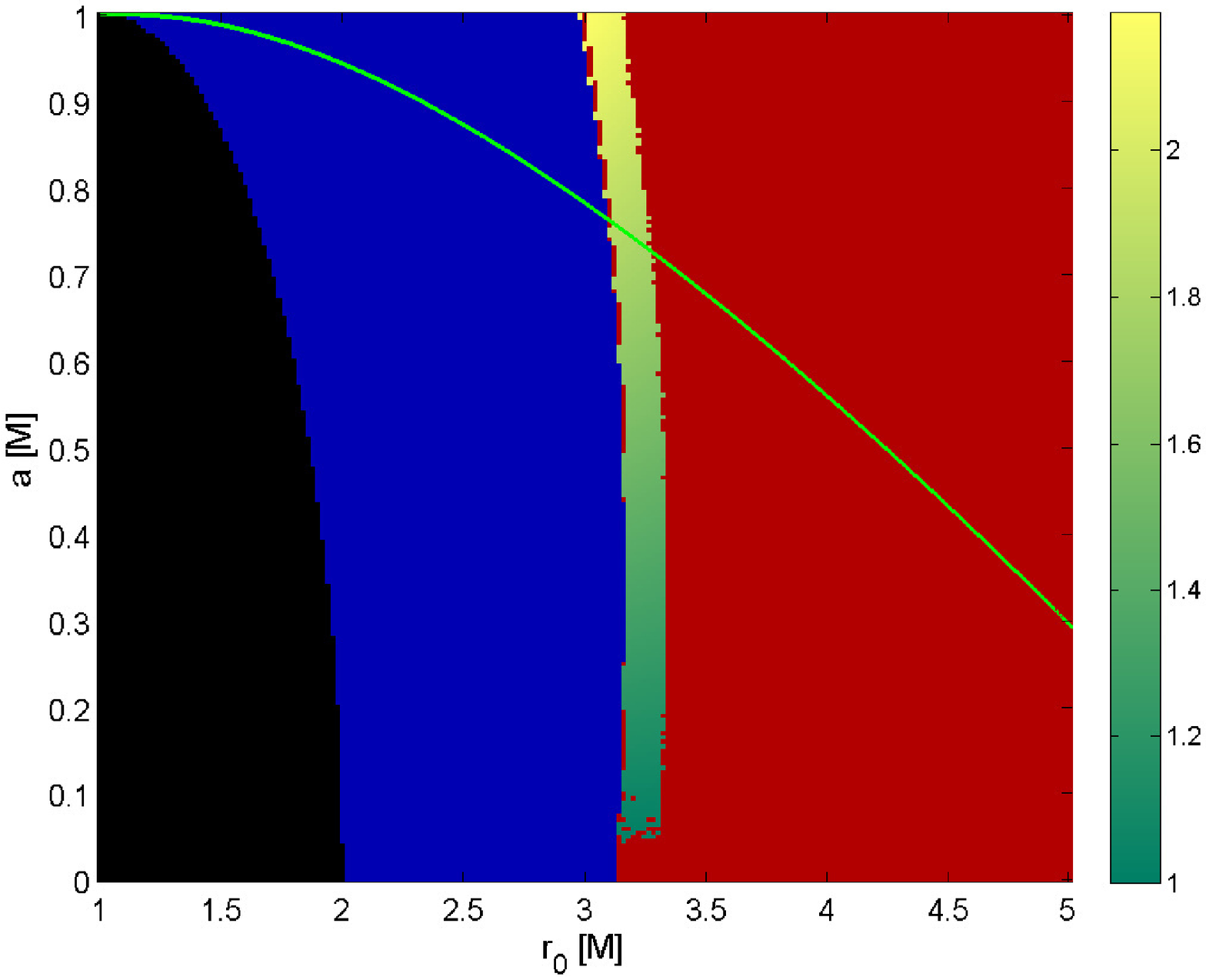}
\includegraphics[scale=.45]{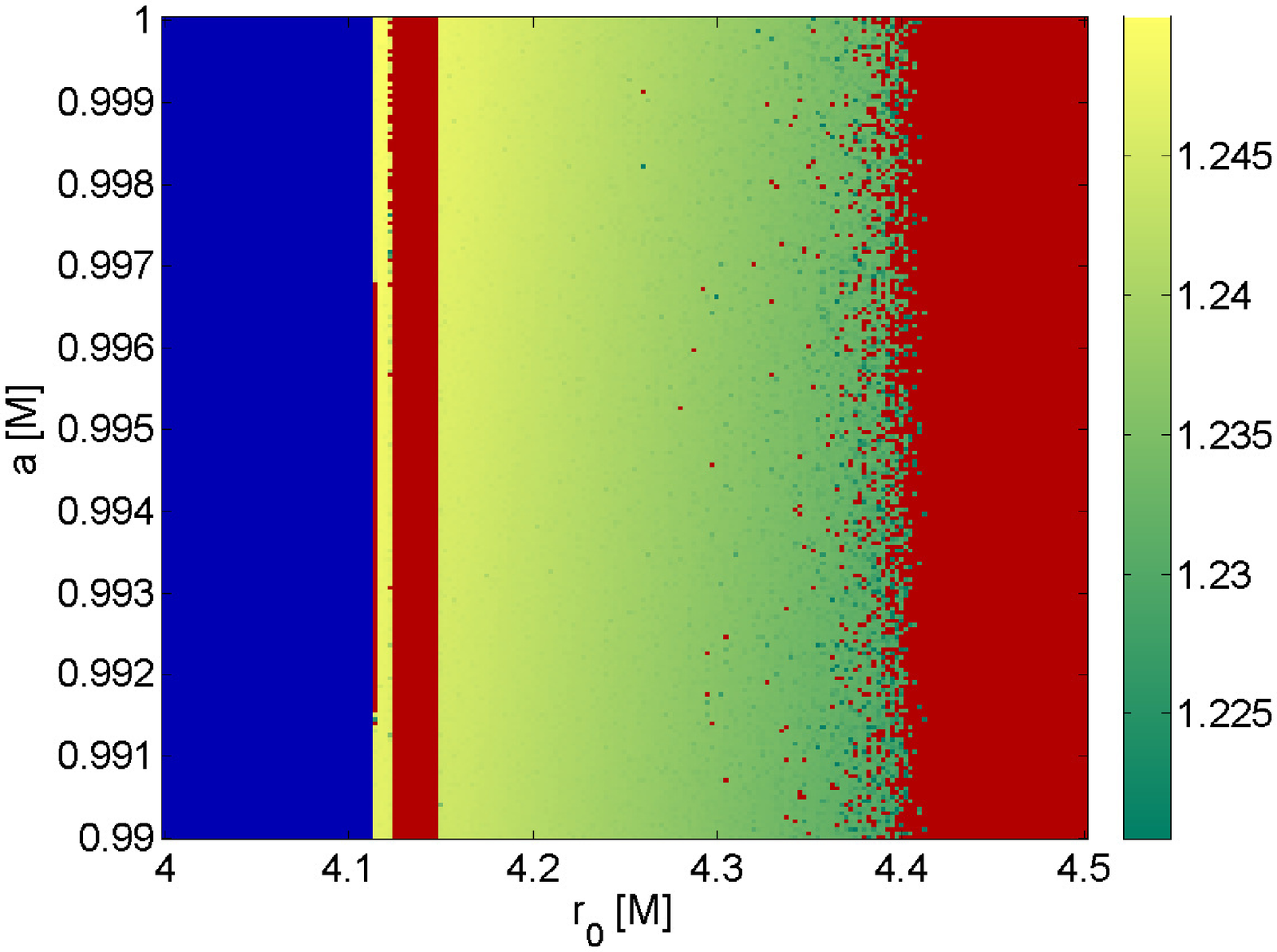}
\caption{Distribution of the final values of Lorentz factor $\gamma$ in the escape zone is encoded by the color scale. The left panel shows the dominant spin dependence of $\gamma$ (demonstrated on the escape zone with $qB=-4$), while the right panel illustrates less prominent dependence on the initial radius (the escape zone with $qB=-1.5$ is shown here). Red color denotes stable (nonescaping) zones ,and blue denotes the region of unstable plunging orbits. Black is the horizon.}
\label{escape_color}
\end{figure*}

In Fig.~\ref{escape_color} we show the distribution of final values of $\gamma$ in the escape zone using the color scale to distinguish particles with different $\gamma$. We observe that acceleration of the escaping particles systematically changes with the spin, as well as with the ionization radius, as suggested by the analysis in Sec.~\ref{ionization}. In particular, in the left panel of Fig.~\ref{escape_color} we demonstrate the spin dependence of the final $\gamma$ and confirm that the highest allowed spin ($a=1$ for this particular case of $qB=-4$) leads to the highest possible acceleration. By choosing a smaller range of spin values, we also notice the radial dependence of final $\gamma$, which is generally less prominent than the former (right panel of Fig.~\ref{escape_color}), and we confirm that lower ionization radius generally leads to higher acceleration. We conclude that for given $qB$, the maximally accelerated particle (with $\gamma_{\rm max}$) corresponds to the highest allowed spin and the lowest possible ionization radius, i.e., in the presented basin-boundary plots it is always located in the upper left corner of the escape zone.  

\begin{figure*}[ht]
\centering
\includegraphics[scale=.45]{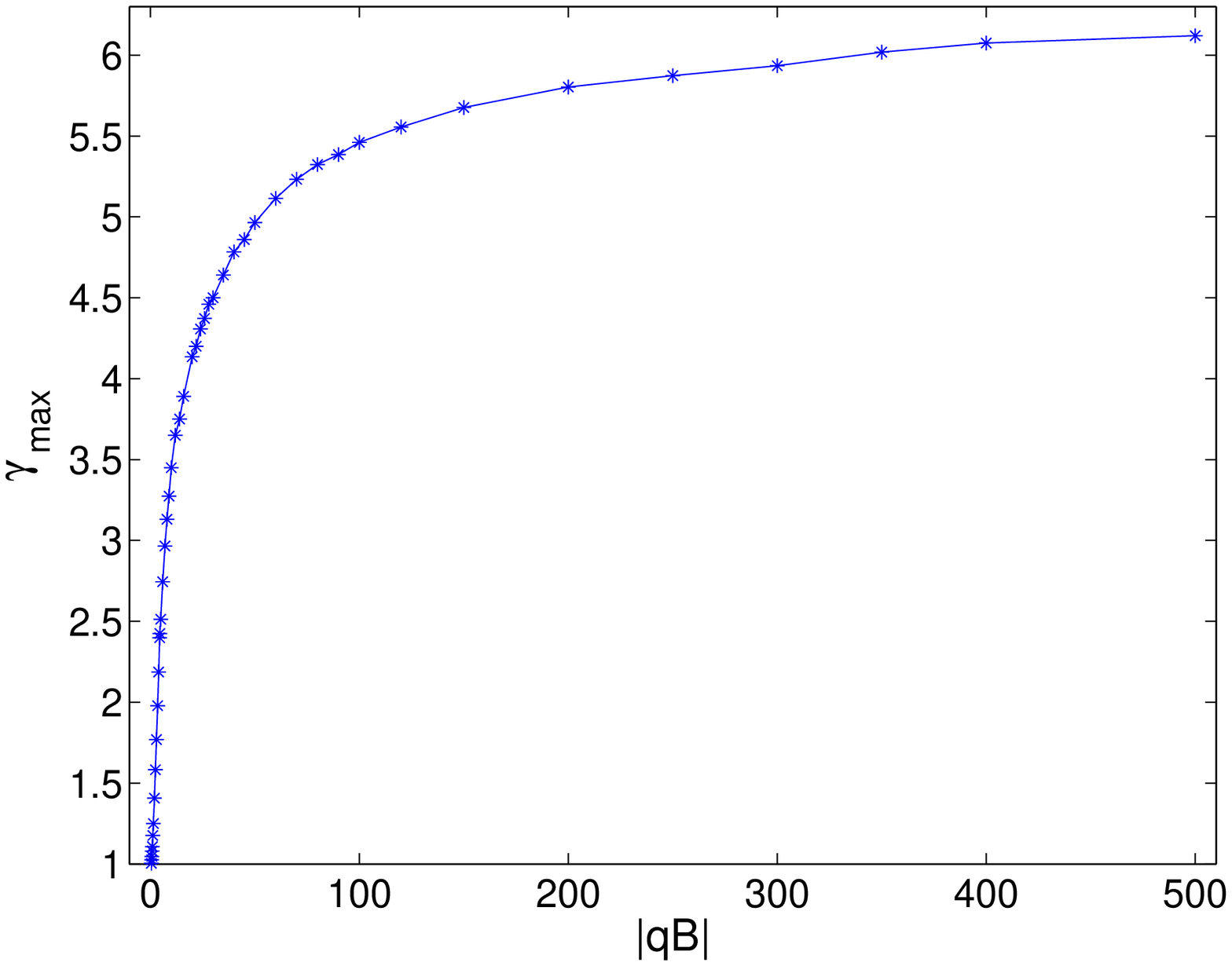}
\includegraphics[scale=.45]{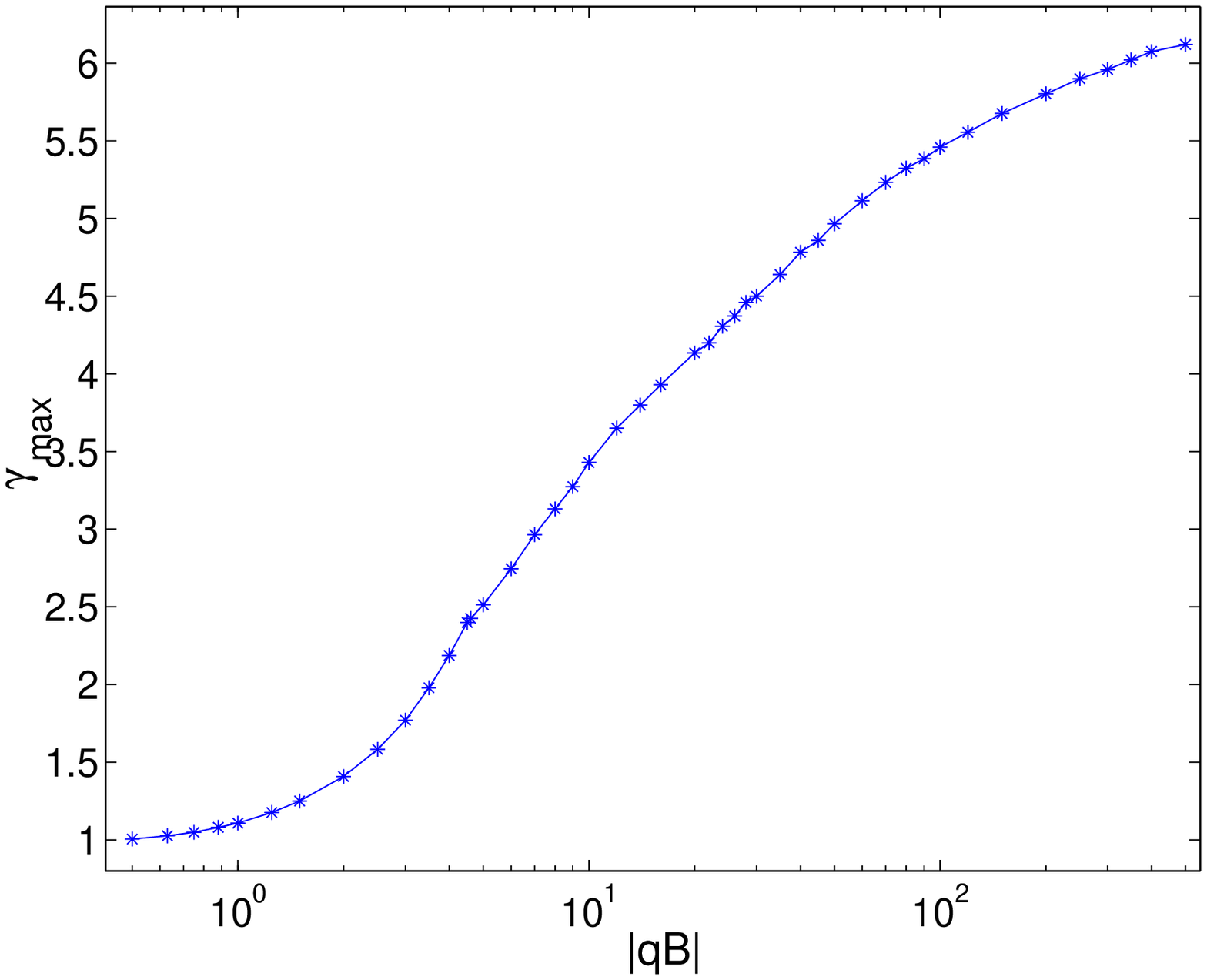}
\caption{Final Lorentz factor $\gamma_{\rm max}$ of the maximally accelerated escaping orbit is shown as a function of the magnitude of magnetization parameter $|qB|$. The both panels only differ in the scaling of $|qB|$ which is linear in the left panel and logarithmic in the right panel.}
\label{gammax}
\end{figure*}

In the following we discuss the dependence of $\gamma_{\rm max}$ on the value of $qB$. For each $qB$ we iteratively integrate trajectories in the relevant segment of the escape zone in order to localize the maximally accelerated   particle and obtain the value of $\gamma_{\rm max}$ with sufficient precision. In Fig.~\ref{gammax} we observe that at first $\gamma_{\rm max}$ rises rapidly with $|qB|$; however, then the growth slows down, and eventually the value seems to saturate at $\gamma_{\rm max}\approx6$. In the right panel of Fig.~\ref{gammax} we present the same data using the logarithmic scale  for $|qB|$, which highlights the sudden change of the shape of the function at $|qB|\approx 4.5$, reflecting the detachment of the escape zone from the extremal spin $a=1$. Higher values of $|qB|$ require gradually lower spins to allow the escape of ionized particles as shown in Fig.~\ref{escape_full}. On the other hand, higher values of spin generally lead to more accelerated escaping orbits as observed in Fig.~\ref{escape_color}. These two mutually opposing tendencies cause the apparent slowdown in the increase of $\gamma_{\rm max}$ above the critical value of $|qB|\approx4.5$. 

\begin{figure*}[ht]
\centering
\includegraphics[scale=.45]{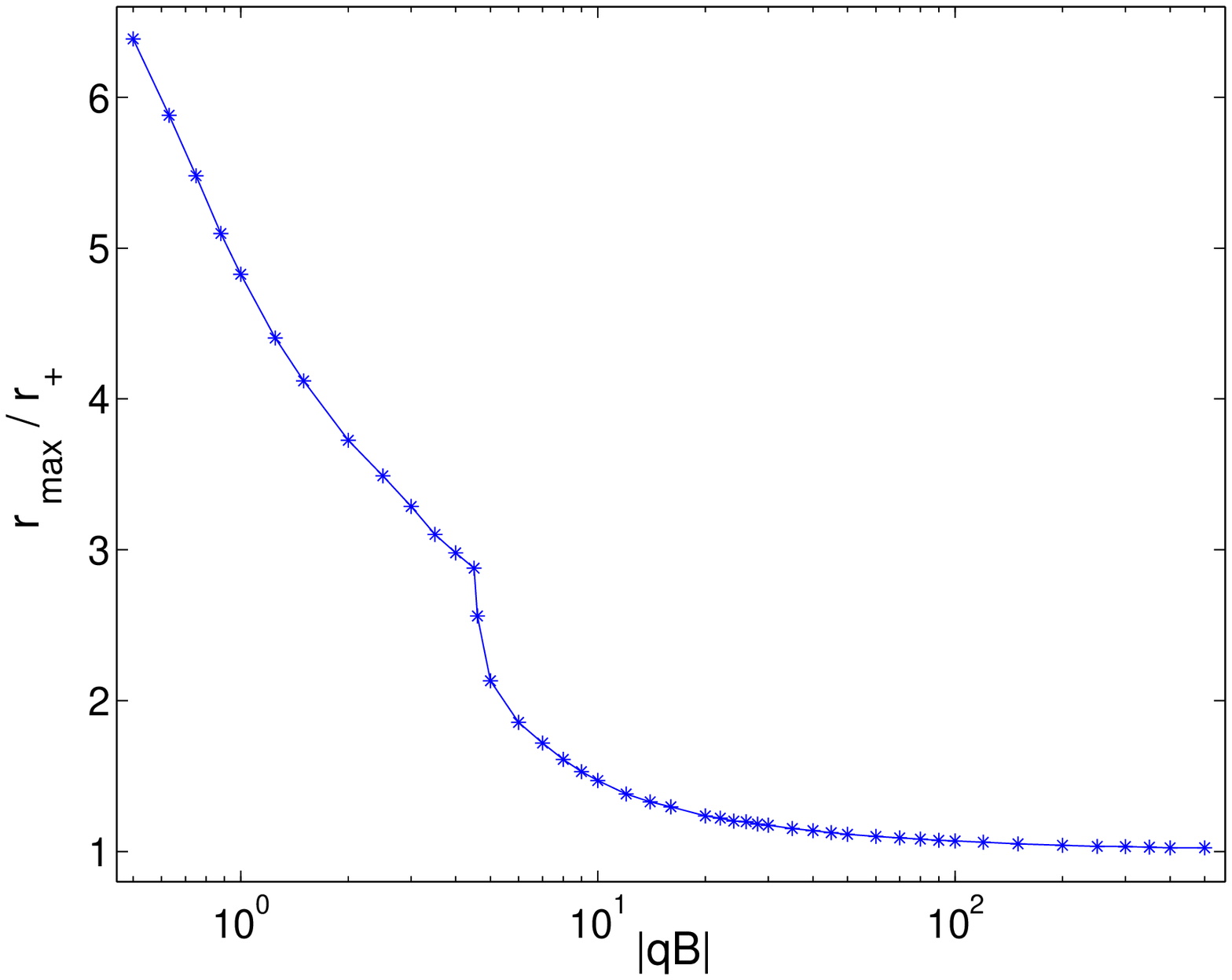}
\includegraphics[scale=.45]{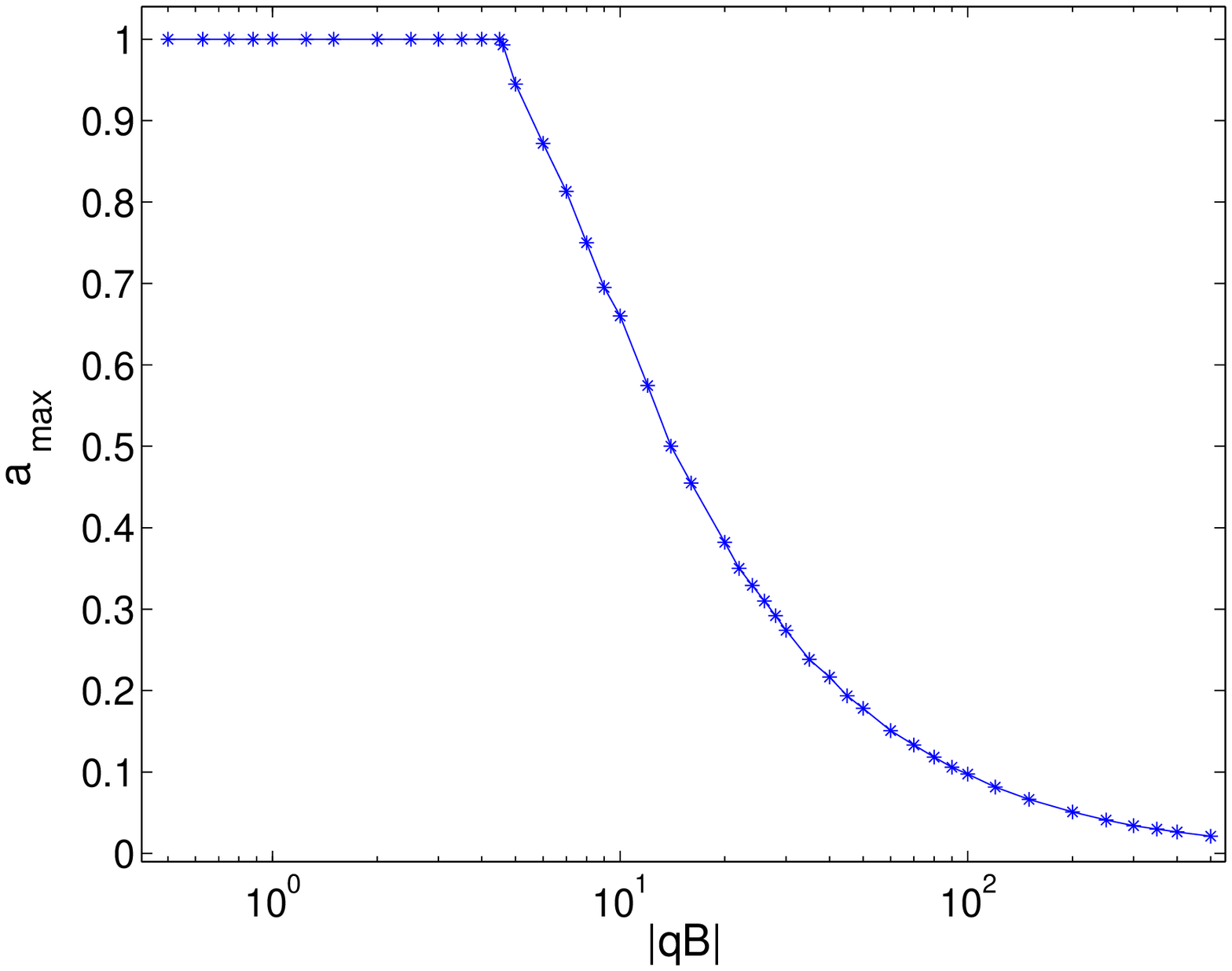}
\caption{Left panel: ratio of the ionization radius of the maximally accelerated particle $r_{\rm max}$ to the radius of the outer horizon $r_+$ as a function of the magnitude of magnetization parameter $|qB|$ (in logarithmic scale). Right panel: analogous dependence of the spin value $a_{\rm max}$ corresponding to the most accelerated particle.}
\label{armax}
\end{figure*}

Increasing the magnetization parameter $|qB|$ shifts the whole escape zone closer to the horizon (as seen in Fig.~\ref{escape_full}), and as a result, the initial radius of the maximally accelerated escaping trajectory $r_{\rm max}$ also gradually approaches the horizon as $|qB|$ increases (see the left panel of Fig.~\ref{armax}). Especially for $|qB|\gtrsim 100$, the value of $r_{\rm max}$ becomes so close to $r_{+}$ that this tendency leads to  growing computational difficulties in tracing the escape zone, as the Boyer--Lindquist coordinates diverge when the horizon is approached. The right panel of Fig.~\ref{armax} shows the value of spin $a_{\rm max}$ corresponding to the most accelerated particle. Above the threshold $|qB|\approx4.5$ the value of $a_{\rm max}$ falls steeply and approaches the static limit $a=0$, which it can never reach, though.

We conclude that the maximal acceleration that can be achieved by charging the Keplerian particle within our model is $\gamma_{\rm max}\approx6$, and it is realized in the case of high magnetization parameter, $|qB|>100$, by particles charged near the horizon of a slowly spinning black hole, $a<0.1$. For extremal black holes, $a=1$, the maximal acceleration is $\gamma_{\rm max}\approx2.5$, corresponding to $qB\approx-4.5$. In accordance with asymptotic behavior of the effective potential (see Eq.~ \ref{asym}), we observe that $\gamma_{\rm max}$ increases with $|qB|$ and within the escape zone it grows with spin $a$ and decreases with $r_0$. However, higher values of $|qB|$ only allow the formation of the escape zone for gradually decreasing $a$, and these opposing trends eventually compensate and let $\gamma_{\rm max}$ saturate at $\gamma_{\rm max}\approx6$.

\section{Discussion}
\label{discussion}

We have numerically evolved particles escaping from the magnetized accretion disk owing to the charging process occurring at inner radii of the disk. The magnetic field is known to act as a nonintegrable perturbation inducing deterministic chaos to the charged particle dynamics. However, in the given setup, the field itself never initiates the motion off the equatorial plane, which can be seen directly from the equations of motion, as well as from the general symmetry considerations. The actual impulse to leave the plane necessarily comes from a different source. In the real accretion disk, this minute yet essential dynamical fluctuation of the orbit in the direction perpendicular to the equatorial plane may be plausibly attributed to the interaction with the nearby matter. In the numerical model of the system  the role of this fluctuation is played by the integration error, which is always present, although its value may be stabilized by the proper choice of the integration scheme \citep{kopacek14b}. In particular, we manage to keep the relative error at the end of each integration below $10^{-10}$ (and the relative local truncation error in every iteration remains at the level $\approx 10^{-14}$).
 
The crucial aspect that draws the distinction between escaping and stable oscillating orbits is the response of the trajectory to the fluctuation. If the particular trajectory is located in the chaotic domain of the phase space, then, due to its sensitivity to initial conditions and exponential growth of any deviation, it will escape if allowed by the topology of the effective potential. On the other hand, the regular trajectory does not further promote the infinitesimal disturbance and remains oscillating around the equatorial plane even if the effective potential allows the escape. 
In the agreement with the general results of chaos theory \citep[see, e.g., ][]{lichtenberg92, contopoulos02}, we indeed observe that some parts of the phase space might be densely populated by both the chaotic escaping trajectories and  the regular oscillating orbits (which is well seen especially in magnified views of the escape zones in Figs.~\ref{escape_odpojeni} and \ref{escape_tail}). Self-similar patterns are typically found in the vicinity of prominent resonances of fundamental frequencies of the system, and we detect them close to the inner edge of the escape zone (Fig.~\ref{escape_detail2}). On the other hand, some portions of the phase space are dominated by regular motion (``islands of stability"), while some portions are mostly chaotic (``chaotic sea"). Nevertheless, even in the chaotic sea we may find small islands of stability (well seen in the right panel of Fig.~\ref{escape_color}, where red-colored stable trajectories are spread over the interior of the escape zone).

Initial orbits of neutral particles were chosen as circular co-rotating geodesics (above the ISCO) and freely inspiraling geodesics (below the ISCO), i.e., they were confined to the equatorial plane ($\theta=\pi/2$). This choice was made for two main reasons. First, the analysis has an obvious astrophysical motivation, and the given initial setup maintains the connection of our model with the standard theory of thin accretion disks. Second, such a choice allows us to study the role that spin and magnetic field play in the collimation and acceleration of the outflow in the especially clear form. Particles on equatorial orbits have a zero velocity component in the direction normal to the disk ($v_{\perp}=0$), and even the charging process itself does not directly produce any (no kick is supposed here). Compared to the scenarios with the kick (i.e., with an ad hoc introduction of $v_{\perp}$) or initially nonequatorial orbits that already have nonzero $v_{\perp}$, the escape in our setup is less probable. We intentionally focused on this case to investigate whether the production of the outflow is possible even under these unfavorable conditions.

Without a preferred direction given by $v_{\perp}$, the orientation of the outflow (i.e., whether the particle escapes upward or rather downward from the equatorial plane) is random in the sense that it depends on the direction of the fluctuation. In the numerical representation of the system we confirm this by observing that the orientation of the escaping trajectory might change by switching the integration scheme or even by changing some of the parameters of the integration routine that affect the step size and the treatment of the local truncation error. In the astrophysical system of accreting black holes we expect the turbulent motion of nearby plasma to disturb the trajectory of charged particles, although the negligible change of the momentum accompanying the photoionization would also suffice to initiate the escape. Due to symmetry of the system, the orientation of a particular escaping trajectory becomes stochastic in a statistical sense, with the equal probability of escaping upward or downward. From the ensemble of charged particles we expect the symmetrical outflows of the both orientations to be formed.

The employed model of the black hole magnetosphere was described by the electromagnetic test field. The test-field solutions are actually a limit of exact solutions for magnetized black holes that have been linearized in electromagnetic terms of the imposed Maxwell fields owing to external sources. Nevertheless, the exact electro-vacuum solution of magnetized Kerr--Newman black holes \citep{ernst76,garcia85} can be treated, in principle, by similar approaches to that adopted in the present paper \citep{karas91}. Nonetheless, the structure of these exact solutions has not yet been understood in the full generality, as they are often asymptotically curved and contain singularities outside the event horizon of the black hole; moreover, nonaxisymmetric exact solutions must be also nonstationary. However, for most astrophysically relevant situations, including the model discussed here, the linearized test-field description remains adequate.

The analysis was performed in a dimensionless units where all quantities are geometrized and scaled by the rest mass of the black hole $M$. To retrieve their values in SI, we express the strength of the magnetic field as follows:
 \begin{equation}
 \label{magfield}
B_{\rm SI}=\frac{qBc_{\rm SI}}{q_{\rm SI}\left(\frac{M}{M_{\odot}}\right)1472\;\rm{m}},
\end{equation}
where the quantities without subscript ${\rm SI}$ are dimensionless and the length $1472\;\rm{m}$ is the solar mass in geometrized units, $M_{\odot}=1472\:\rm{m}$. As a typical value of magnetization parameter $qB$ for which the escape zone forms also for high spin values, we insert $qB=-1$.

To explore the upper limit of the specific charge $q_{\rm SI}$, we set the value for electrons, i.e., $q_{\rm SI}=-1.76\times10^{11}\; \rm{C}\,\rm{kg}^{-1}$, for which we find that for a stellar mass black hole of $M=10\;M_{\odot}$ the corresponding magnetic field reads $B_{\rm SI}=1.16\times10^{-7}\;\rm{T}$. Such a value is consistent with nonthermal filaments observed in the Galactic Center \citep{larosa04,ferr10}.

Nevertheless, a typical specific charge obtained by a dust grain  would be considerably lower. If we consider the supermassive black hole in the center of the giant elliptical galaxy M87 with mass $M = 7.22^{+0.34}_{-0.40} \times 10^9\; M_{\odot}$ \citep{oldham16}, we may use the estimates of the field strength derived from the Event Horizon Telescope data by \citet{kino14,kino15}. The field at the base of the jet in M87 is constrained as $5\times10^{-3}\; \rm{T}\leq \emph{B} \leq 1.24\times 10^{-2}\; \rm{T}$ \citep{kino15}, while at a distance of the scale of 10 Schwarzschild radii from the source they get $10^{-4}\; \rm{T}\leq \emph{B} \leq 1.5\times 10^{-3} \;\rm{T}$ \citep{kino14}. Inserting these values into Eq.~(\ref{magfield}), we obtain a specific charge of the particle in the range $-0.3 \; \rm{C}\,\rm{kg}^{-1} \leq \emph{q}_{\rm SI} \leq -2\times 10^{-3} \; \rm{C}\,\rm{kg}^{-1}$, which is plausible for the charged dust grains. For the microquasar GRS $1915\!+\!105$, which is powered by the stellar mass black hole with $M = 12.4^{+2.0}_{-1.6}\; M_{\odot}$ \citep{reid14}, the magnetic field strength in the wind zone was constrained by the upper limit in the range $0.1\; \rm{T}\leq \emph{B} \leq 10\; \rm{T}$ \citep{miller16}. For these values Eq.~(\ref{magfield}) yields the specific charge  $-7.1 \times 10^{-4}\; \rm{C}\,\rm{kg}^{-1} \leq \emph{q}_{\rm SI} \leq -5.3\times 10^{-6} \; \rm{C}\,\rm{kg}^{-1}$, which suggests that ionization to lower charges (compared to the supermassive black hole in M87) is sufficient for the escape of particles from GRS $1915\!+\!105$ system.

Although we do not reduce our analysis to describe a particular  astrophysical object, it is apparent that the given scenario is astrophysically relevant and generally compatible with the conditions encountered in accreting black hole systems.

\section{Conclusions}
\label{conclusions}
We have analyzed the effect of outflow of matter from the inner part of the magnetized accretion disk due to charging of initially neutral accreted dust. In particular, we have shown that a rather simple setup is sufficient to model an outflow even in the single particle approximation, where magnetohydrodynamic terms are neglected. Indeed, it appears that essential and sufficient ingredients to allow the matter to escape are the rotation of the central black hole, $a\neq0$, and the presence of uniform magnetic field $B$. We assumed a basic accretion scenario in which the matter slowly sinks between circular Keplerian orbits (and becomes free-falling below the ISCO) in the equatorial plane, until it reaches charging radius $r_0$, where it obtains nonzero charge $q$. In general, the charge may stabilize the plunging orbit, as well as possibly turning the stable orbit into plunging. From the analysis of the effective potential we found that only negatively charged particles could possibly escape the attraction of a black hole with parallel orientation of spin and magnetic field.

A numerical survey revealed that escaping trajectories are actually realized only for $qB\lessapprox -0.5$. Until $qB\approx-4.5$, the escape zone stretches from extremal spin $a=1$ to gradually lower spin values. However, for $qB\lessapprox-4.5$ it disconnects from the $a=1$ limit and the whole escape zone is gradually compressed to lower spins and smaller radii as the $|qB|$ increases. The maximal Lorentz factor $\gamma_{\rm max}$ that can be achieved by escaping particles was calculated. To some surprise, it seems to saturate at $\gamma_{\rm max}\approx6$, which is attained with $|qB|\gtrapprox 100$ for $a\lessapprox 0.1$. Maximally spinning black holes accelerate the escaping matter up to $\gamma_{\rm max}\approx2.5$ for $qB\approx-4.5$.

The presence of the magnetic field was crucial for the outflow for two complementary reasons. First, it changes the topology of the effective potential, so that it makes the asymptotic region in principle accessible ("potential valleys" connecting the equatorial plane with the asymptotic region are formed). The necessary condition for the existence of escaping trajectories is thus fulfilled. Another essential ingredient is, however, the impact of the field on the dynamic regime of charged particles. The magnetic field in the vicinity of a rotating black hole is known to trigger chaotic motion. And, as we suggest, only chaotic particles may escape from the circular orbits in the accretion disk, while regular ones remain oscillating in the equatorial plane even if the charge gives them enough energy to escape. This also explains why the escape zone may only form close to the black hole, where the system is sufficiently nonintegrable. Farther from the horizon, the spacetime becomes almost flat and magnetic field almost uniform, which leads to regular motion of charged particles. 

Even though we considered a basic model, the analysis provides deep insight into the undisputed and crucial role of ordered magnetic field for the formation of the astrophysical jet. Most importantly, it shows that even in the particle approximation where magnetohydrodynamic terms are neglected,  the outflow may be formed and attain relativistic velocities.

\acknowledgements
Authors acknowledge the support from the COST CZ program of the Czech Ministry of Education (project LD15061). O.K. is grateful for support from the Czech Science Foundation (GA\v{C}R-17-06962Y). V.K. thanks the Czech Science Foundation for grant GA\v{C}R-14-37086G (``Albert Einstein Center for Gravitation and Astrophysics"). Discussions with Martin Kolo\v{s} are highly appreciated.


\begin{thebibliography}{}
\bibitem[Al Zahrani et al.(2014)]{alzahrani14}Al Zahrani, A.~M. 2014, \prd, 90, 044012
\bibitem[Al Zahrani et al.(2013)]{alzahrani13}Al Zahrani, A.~M., Frolov, V.~P., \& Shoom, A.~A. 2013, \prd, 87, 084043
\bibitem[Babar et al.(2016)]{babar16}Babar, G.~Z., Jamil, M., \& Lim, Y.-K. 2016, Int. J. Mod. Phys. D, 25, 1650024
\bibitem[Balbus \& Hawley(1998)]{balbus98}Balbus, S.~A., \& Hawley, J.~F. 1998, Rev. Mod. Phys., 70, 1-53
\bibitem[Bardeen et al.(1972)]{bardeen72}Bardeen, J. M., Press, W. H., \& Teukolsky, S. A. 1972, \apj, 178, 347-370
\bibitem[Beckwith et al.(2008)]{beckwith08}Beckwith, K., Hawley, J.~F., \& Krolik, J.~H. 2008, \apj, 678, 1180-1199
\bibitem[Blandford \& Payne(1982)]{blandford82}Blandford, R.~D., \& Payne D.~G. 1982, \mnras, 199, 883-903
\bibitem[Blandford \& Znajek(1977)]{blandford77}Blandford, R.~D., \& Znajek, R.~L. 1977, \mnras, 179, 433-456 
\bibitem[Contopoulos(2002)]{contopoulos02}Contopoulos, G. 2002, Order and chaos in dynamical Astronomy (Springer)
\bibitem[Ernst \& Wild(1976)]{ernst76}Ernst, F.~J., \& Wild, W.~J. 1976, J. Math. Phys., 17, 182-184
\bibitem[Ferri\`{e}re(2010)]{ferr10}Ferri\`{e}re, K. 2010,  Astron. Nachr., 331, 27-33
\bibitem[Frolov \& Shoom(2010)]{frolov10}Frolov, V.~P., \& Shoom, A.~A. 2010, \prd, 82, 084034
\bibitem[Garc\'{\i}a D\'{\i}az(1985)]{garcia85}Garc\'{\i}a D\'{\i}az, A. 1985, J. Math. Phys., 26, 155-156
\bibitem[Gold et al.(2017)]{gold17}Gold, R., McKinney, J.~C., Johnson, M.~D., \& Doeleman, S.~S. 2017, \apj, 837, 180
\bibitem[Huang et al.(2015)]{huang15}Huang, Q., Chen, J., \& Wang, Y. 2015, Int. J. Mod. Phys. D, 24, 1550054 
\bibitem[Hussain et al.(2014)]{hussain14}Hussain, S., Hussain, I., \& Jamil, M. 2014, Eur. Phys. J. C, 74:3210 
\bibitem[Karas et al.(2014)]{karas14}Karas, V., Kop\'{a}\v{c}ek, O., Kunneriath, D., \& Hamersk\'{y}, J. 2014, Acta Polytechnica, 54, 398-413
\bibitem[Karas et al.(2012)]{karas12}Karas, V., Kop\'{a}\v{c}ek, O., \& Kunneriath, D. 2012, Classical Quant. Grav., 29, 035010
\bibitem[Karas \& Kop\'{a}\v{c}ek(2009)]{karas09}Karas, V., \& Kop\'{a}\v{c}ek, O. 2009, Classical Quant. Grav., 26, 025004
\bibitem[Karas \& Vokrouhlick\'{y}(1991)]{karas91}Karas, V., \& Vokrouhlick\'{y}, D. 1991, J. Math. Phys., 32, 714-716 
\bibitem[Kino et al.(2015)]{kino15}Kino, M., Takahara, F., Hada, K., Akiyama, K., Nagai, H., \& Sohn, B.~W. 2015, \apj, 803, 30
\bibitem[Kino et al.(2014)]{kino14}Kino, M., Takahara, F., Hada, K., \& Doi, A. 2014, \apj, 786, 5
\bibitem[Kolo\v{s} et al.(2015)]{kolos15}Kolo{\v s}, M., Stuchl{\'{\i}}k, Z., \& Tursunov, A. 2015, Classical Quant. Grav., 32, 165009
\bibitem[Kop\'{a}\v{c}ek et al.(2014)]{kopacek14b}Kop\'{a}\v{c}ek, O., Karas, V., Kov\'{a}\v{r}, J., \& Stuchl\'{i}k, Z. 2014, in Proc. of RAGtime 10-13, eds. Z. Stuchl\'{i}k, G. T\"{o}r\"{o}k, T. Pech\'{a}\v{c}ek (Opava: Silesian Univ.), pp. 123-132, arXiv:1601.01262
\bibitem[Kop\'{a}\v{c}ek \& Karas(2014)]{kopacek14}Kop\'{a}\v{c}ek, O., \& Karas, V. 2014, \apj, 787, 117
\bibitem[Kop\'{a}\v{c}ek et al.(2010)]{kopacek10}Kop\'{a}\v{c}ek, O., Karas, V., Kov\'{a}\v{r}, J., \& Stuchl\'{i}k, Z. 2010, \apj, 722, 1240
\bibitem[Kop\'{a}\v{c}ek et al.(2010b)]{kopacek10b}Kop\'{a}\v{c}ek, O., Kov\'{a}\v{r}, J., Karas, V., \& Stuchl\'{i}k, Z. 2010b, in Proc. Mathematics and Astronomy: A Joint Long Journey, eds. M. de Le\'{o}n, D. M. de Diego \& R. M. Ros, (Melville, NY: Springer), pp. 278-287 
\bibitem[Kov\'{a}\v{r} et al.(2013)]{kovar13}Kov\'{a}\v{r}, J., Kop\'{a}\v{c}ek, O., Karas, V., \& Kojima, Y. 2013, Classical Quant. Grav., 30, 025010
\bibitem[Kov\'{a}\v{r} et al.(2010)]{kovar10}Kov\'{a}\v{r}, J., Kop\'{a}\v{c}ek, O., Karas, V., \& Stuchl\'{i}k, Z. 2010, Classical Quant. Grav., 27, 135006
\bibitem[Kov\'{a}\v{r} et al.(2008)]{kovar08}Kov\'{a}\v{r}, J., Stuchl\'{i}k, Z., \& Karas, V. 2008, Classical Quant. Grav., 25, 095011
\bibitem[LaRosa et al.(2004)]{larosa04}LaRosa, T.~N., Nord, M.~E., Lazio, T.~J.~W., \& Kassim, N.~E. 2004, \apj, 607, 302-308
\bibitem[Lichtenberg \& Lieberman(1992)]{lichtenberg92}Lichtenberg, A. J., \& Lieberman, M. A. 1992, Regular and chaotic dynamics (New York: Springer) 
\bibitem[Lukes-Gerakopoulos \& Kop\'{a}\v{c}ek(2018)]{lukes18}Lukes-Gerakopoulos, G., \& Kop\'{a}\v{c}ek, O. 2018, Int. J. Mod. Phys. D, 27, 1850010
\bibitem[Marwan et al.(2007)]{marwan07}Marwan, N., Carmen Romano, M., Thiel M., \& Kurths J. 2007, Phys. Rep., 438, 237  
\bibitem[McKinney et al.(2013)]{mckinney13}McKinney, J.~C., Tchekhovskoy, A., \& Blandford, R.~D. 2013, Science, 339, 49
\bibitem[Miller et al.(2016)]{miller16}Miller, J.~M., Raymond, J., Fabian, A.~C., Gallo, E., Kaastra, J., Kallman, T., King, A.~L., Proga, D., Reynolds, C.~S., \& Zoghbi, A. 2016, \apjl, 821, L9
\bibitem[Misner et al.(1973)]{mtw}Misner, C. W., Thorne, K. S., \& Wheeler, J. A. 1973 Gravitation (San Francisco: Freeman)
\bibitem[Novikov \& Thorne(1973)]{novikov73}Novikov, I. D., \& Thorne, K. S. 1973, in Black Holes - Les Astres Occlus, eds. C. De Witt and B. S. De Witt (New York: Gordon \& Breach), pp. 343-450
\bibitem[Oldham \& Auger(2016)]{oldham16}Oldham, L.~J., \& Auger, M.~W. 2016, \mnras, 457, 421-439
\bibitem[Penna et al.(2010)]{penna10}Penna, R. F., McKinney, J. C., Narayan, R., Tchekhovskoy, A., Shafee, R., \& McClintock, J. E. 2010, \mnras, 408, 752
\bibitem[Reid et al.(2014)]{reid14}Reid, M.~J., McClintock, J.~E., Steiner J.~F., Steeghs, D., Remillard, R.~A., Dhawan, \& Narayan, R. 2014, \apj, 796, 2
\bibitem[Sadowski(2016)]{sadowski16}Sadowski, A. 2016, \mnras, 459, 4397-4407
\bibitem[Semer\'{a}k \& Sukov\'{a}(2012)]{semerak12}Semer\'{a}k, O., \&  Sukov\'{a}, P. 2012, \mnras, 425, 2455
\bibitem[Shiose et al.(2014)]{shiose14}Shiose, R., Kimura, M., \& Chiba, T. 2014, \prd, 90, 124016
\bibitem[Skokos(2010)]{skokos10}Skokos, Ch. 2010, in Dynamics of Small Solar System Bodies and Exoplanets, eds. J. Souchay and R. Dvorak, Lect.\ Notes Phys.\ 790 (Springer: Berlin), pp.~63-135
\bibitem[Stuchl\'{i}k \& Kolo\v{s}(2016)]{stuchlik16}Stuchl\'{i}k, Z., \& Kolo\v{s}, M. 2016, Eur. Phys. J. C, 76, 32
\bibitem[Sukov\'{a} \& Janiuk(2016)]{sukova16}Sukov{\'a}, P., \& Janiuk, A. 2016, Astron. Astrophys., 591, A77 
\bibitem[Tchekhovskoy(2015)]{tchekhovskoy15}Tchekhovskoy, A. 2015, Astrophys. Space Sc. L., 414, 45-82
\bibitem[T\'{e}l \& Gruiz(2006)]{tel06}T{\'e}l, T. \& Gruiz, M. 2006, Chaotic Dynamics: An Introduction Based on Classical Mechanics (Cambridge University Press) 
\bibitem[Tursunov et al.(2016)]{tursunov16}Tursunov, A., Stuchl{\'{\i}}k, Z., \& Kolo{\v s}, M. 2016, \prd, 93, 084012
\bibitem[Wald(1974)]{wald74}Wald, R. M. 1974, \prd, 10, 1680
\bibitem[Wilkins(1972)]{wilkins72}Wilkins, D. C. 1972, \prd, 5, 814-822
\bibitem[Witzany et al.(2015)]{witzany15}Witzany, V., Semer{\'a}k, O., \& Sukov{\'a}, P. 2015, \mnras, 451, 1770-1794
\end{thebibliography}
\end{document}